\documentclass[journal]{IEEEtran}

\usepackage{amsmath, amsthm, amssymb, mathtools, mathrsfs, bbm}
\usepackage{relsize, paralist, balance, hyperref, cite, xcolor}
\usepackage{graphicx, subfigure, caption}

\newtheorem{theorem}{Theorem}%[section]
\newtheorem{lemma}{Lemma}
\newtheorem{proposition}[theorem]{Proposition}
\newtheorem{corollary}[theorem]{Corollary}

\newtheorem{remark}{Remark}
\newtheorem*{example}{Example}

\newcommand{ \bs }[1]{ \boldsymbol{#1} }
\newcommand{ \N }{ \mathbb{N} }
\newcommand{ \Z }{ \mathbb{Z} }
\newcommand{ \A }{ \mathcal A }
\newcommand{ \B }{ \mathcal B }
\newcommand{ \C }{ \bs{C} }
\newcommand{ \myb }{ \textnormal{b} }
\newcommand{ \myx }{ \bs{x} }
\newcommand{ \myS }{ \bs{S} }
\newcommand{ \mySSr }{ \bs{S} }
\newcommand{ \mySr }{ S }
\newcommand{ \mysr }{ \sigma }
\newcommand{ \lmin }{ \inf L }
\newcommand{ \lmax }{ \sup L }
\newcommand{ \weight }{ \operatorname{wt} }
\newcommand{ \run }{ \operatorname{run} }
\newcommand{ \myqed }{ \hfill $\blacktriangle$ }

\begin{document}

\title{Asymptotic Behavior and Typicality Properties of Runlength-Limited Sequences}

\author{
        Mladen~Kova\v{c}evi\'c %,~\IEEEmembership{Senior~Member,~IEEE}
				%ORCiD: 0000-0002-2395-7628
				and
				Dejan Vukobratovi\'{c} %,~\IEEEmembership{Senior~Member,~IEEE}
				%ORCiD: 0000-0002-5305-8420
\thanks{Date: December 9, 2021.}
\thanks{The authors are with the Faculty of Technical Sciences,
        University of Novi Sad, 21000 Novi Sad, Serbia
				(emails: \{kmladen, dejanv\}@uns.ac.rs).}
\thanks{This work was supported by the European Union's Horizon 2020
        research and innovation programme under Grant Agreement number 856967,
				and by the Secretariat for Higher Education and Scientific Research
				of the Autonomous Province of Vojvodina through the project ``Visible light
				technologies for indoor sensing, localization and communication in smart buildings''.
				The work was presented in part at the 2021 IEEE Information Theory Workshop (ITW) \cite{kov+vuk}.}
}

\maketitle

\begin{abstract}
This paper studies properties of binary runlength-limited sequences
with additional constraints on their Hamming weight and/or their number
of runs of identical symbols.
An algebraic and a probabilistic (entropic) characterization of the
exponential growth rate of the number of such sequences, i.e., their
information capacity, are obtained by using the methods of multivariate
analytic combinatorics, and properties of the capacity as a function
of its parameters are stated.
The second-order term in the asymptotic expansion of the rate of these
sequences is also given, and the typical values of the relevant
quantities are derived.
Several applications of the results are illustrated, including bounds
on codes for weight-preserving and run-preserving channels (e.g., the
run-preserving insertion-deletion channel), a sphere-packing bound for
channels with sparse error patterns, and the asymptotics of constant-weight
sub-block constrained sequences.
In addition, the asymptotics of a closely related notion---%
\boldmath{$ q $}-ary sequences with fixed Manhattan weight---is briefly
discussed, and an application in coding for molecular timing channels
is illustrated.
\end{abstract}


\section{Introduction}

\IEEEPARstart{I}{n most} data recording and communication systems,
some data sequences are more susceptible to errors than others.
Constrained codes are used for the purpose of avoiding such sequences
and thereby reducing the possibility of an erroneous symbol detection
or a synchronization fault.
Due to their usefulness in designing reliable information storage systems,
constrained codes have found applications in hard disk, non-volatile
memories, optical discs, etc. \cite{immink_book, marcus}, and they are
also projected for usage in future DNA storage systems \cite{immink+cai}.
This paper is devoted to an important class of constrained sequences
called runlength-limited (RLL) sequences, which have been widely studied
and applied in both line coding and error control coding contexts
\cite{immink_book, marcus}.
We note, however, that the methods used in the paper are applicable
to a wider class of constraints;
this will be illustrated on the example of the so-called sub-block
constrained sequences.

The emphasis in the paper is placed on the asymptotic analysis.
By imposing additional constraints on RLL codes, we shall refine the
classical results about the achievable information rates thereof and
obtain more precise asymptotic statements.
In particular, the additional constraints we consider are:
\begin{inparaenum}
\item[i)]
the \emph{constant-weight} constraint, i.e., the requirement that all
the codewords have the same Hamming weight, and
\item[ii)]
the \emph{constant-number-of-runs}%
\footnote{The constant-number-of-runs condition can also be expressed
as a constant-weight condition in a different domain, see
Remark \ref{rem:dk} ahead.}
constraint, i.e., the requirement
that all the codewords have the same number of runs of identical symbols.
\end{inparaenum}
Constant-weight and bounded-weight codes have numerous applications
in communications (see, e.g., \cite{immink+cai, kovacevic2, kurmaev, kurmaev2}
for a study of constant-weight codes in the context of runlength
constraints).
Apart from these, the motivation behind the above-mentioned constraints
that we analyze here is twofold:
\begin{inparaenum}
\item[1)]
on the theoretical side, to quantify precisely the asymptotic behavior
and derive the \emph{typical} values of the relevant quantities in RLL
sequences, and
\item[2)]
on the application side, to exhibit their usefulness in the analysis of
various communication scenarios.
\end{inparaenum}

The main results of the paper regarding the asymptotic properties of
RLL sequences are presented in Section~\ref{sec:rll}.
These include the asymptotic rates of constant-weight and constant-number-of-runs
RLL codes, probabilistic (entropic) characterization of the corresponding
information capacities, properties of these capacities as functions of
their parameters, and the typical values of several relevant quantities
concerning RLL sequences.
In Section~\ref{sec:applications} we describe three examples of communication
scenarios for which the results are relevant:
\begin{inparaenum}
\item[1)]
weight-preserving and run-preserving channels (in particular, the deletion
channel with RLL inputs),
\item[2)]
channels with sparse error patterns (in which the noise sequences, rather
than the information sequences, are constrained), and
\item[3)]
channels with sub-block constrained inputs, which are relevant for, e.g.,
simultaneous information and energy transmission.
\end{inparaenum}
In Section~\ref{sec:manhattan} we discuss the asymptotics of $ q $-ary
sequences with fixed Manhattan weight, objects which are, in a sense,
dual to constant-number-of-runs RLL sequences, and we demonstrate their
application in coding for Manhattan-weight-preserving channels such as
the molecular timing channel.

\subsection*{Notation}

The Hamming weight of a string/sequence $ \bs{x} = x_1 \cdots x_n \in \{0, 1\}^n $
is denoted by $ \weight(\myx) = |\{ i : x_i \neq 0 \}| = \sum_{i=1}^n x_i $.
By a \emph{run} of identical symbols in $ \bs{x} $ we always mean a
maximal run, i.e., a substring of identical symbols that is delimited
on both sides either by a different symbol, or by the beginning/end of
the string $ \bs{x} $.
The number of runs in $ \bs{x} $ is denoted by $ \run(\bs{x}) $; for example,
the string $ 0100011 $ has $ 4 $ runs.
The string consisting of $ \ell $ identical symbols $ a \in \{0, 1\} $ is
denoted by $ a^\ell $.
The symbol $ \oplus $ stands for the XOR operation (addition modulo $ 2 $).
$ H(X) \equiv H(P) = - \sum_{i} p_i \log p_i $ is the Shannon entropy
of a random variable $ X $ having probability distribution $ P = (p_i) $,
and $ \log $ is the base-$ 2 $ logarithm.
In the Bernoulli case, $ P = (p, 1 - p) $, we shall abuse the notation
slightly and write $ H(p) $ for $ H(P) $.
$ \mathbb{N} = \{1, 2, \ldots\} $ is the set of natural numbers.
For a subset $ L \subseteq \N $ and an integer $ s $, we denote by
$ L + s = \{ \ell + s : \ell \in L \} $ the translation of $ L $ by $ s $.
For two non-negative real sequences $ (a_n) $ and $ (b_n) $:
\begin{inparaenum}
\item
$ a_n \sim b_n $ means $ \lim_{n \to \infty} \frac{a_n}{b_n} = 1 $;
\item
$ a_n \gtrsim b_n $ means $ \liminf_{n \to \infty} \frac{a_n}{b_n} \geqslant 1 $;
\item
$ a_n = {\mathcal O}(b_n) $ means $ \limsup_{n \to \infty} \frac{a_n}{b_n} < \infty $;
\item
$ a_n = o(b_n) $ means $ \lim_{n \to \infty} \frac{a_n}{b_n} = 0 $.%
\end{inparaenum}%

\section{Asymptotics of Runlength-Limited Sequences}
\label{sec:rll}

Fix a subset $ L \subseteq \N $, $ |L| \geqslant 2 $.
Let $ \myS_{L}(n) $ denote the set of all sequences of length $ n $ that are
built from blocks in $ \{ 0^\ell : \ell \in L \} $ and $ \{ 1^\ell : \ell \in L \} $
in an alternating manner, meaning that a block of zeros is followed by a
block of ones and vice versa.
In other words, $ \myS_{L}(n) $ is the set of binary sequences of length
$ n $ in which the lengths of all runs of identical symbols belong to $ L $.
These sequences are referred to as runlength-limited (RLL) sequences%
\footnote{In Section \ref{sec:different} we also discuss the more general
case when the lengths of runs of $ 0 $'s and those of $ 1 $'s have possibly
different constraints.}.
Hence, $ L $ is the set of lengths of the allowed constituent blocks,
the smallest length being $ \lmin \in \N $, and the smallest upper
bound on the lengths being $ \lmax \in \N \cup \{\infty\} $.
We shall assume hereafter that $ \gcd L = 1 $.
This condition is equivalent to saying that all but possibly finitely
many elements of $ \N $ can be obtained as non-negative integer combinations
of the elements of $ L $ and, hence, that $ \myS_{L}(n) $ is non-empty
for all $ n \geqslant n_0 $.
If this is not the case, the asymptotic statements we shall give remain
valid, but the condition $ {n \to \infty} $ is then to be understood over
the semigroup generated by $ L $.
E.g., if $ L = \{2, 4\} $, then the RLL sequences defined above are
necessarily of even length.

\begin{remark}
\label{rem:dk}
\textnormal{
By using the transformation $ \bs{x} \mapsto \bs{x'} $ defined by
$ x_i' = x_i \oplus x_{i-1} $, $ i = 1, \ldots, n $, where we understand
that $ x_0 = 0 $, the set of RLL sequences is mapped to the set of sequences
with constrained \emph{runs of zeros} \cite{immink}.
More precisely, any two successive ones in $ \bs{x'} $ are separated by a
run of zeros whose length belongs to the set $ L - 1 = \{ \ell - 1 : \ell \in L \} $.
Therefore, studying these two types of constraints are essentially equivalent
problems.
Note also that
\begin{equation}
\label{eq:dk}
  \weight(\bs{x'}) =
	\begin{cases}
	 \run(\bs{x})     , & \textnormal{if}\;\; x_1 = 1  \\
	 \run(\bs{x}) - 1 , & \textnormal{if}\;\; x_1 = 0 .
	\end{cases}
\end{equation}
We shall refer back to this fact in Section \ref{sec:ronly}.
}

\textnormal{
Binary sequences in which successive ones are separated by runs of zeros
whose lengths are constrained to the set $ \{d, d+1, \ldots, k\} $, for
some $ 0 \leqslant d < k \leqslant \infty $, are called\linebreak
$ (d, k) $-sequences (here by $ k = \infty $ we mean $ \{d, d+1, \ldots\} $).
In our notation, they correspond to the case $ L = \{ d+1, d+2, \ldots, k+1 \} $.
}
\myqed
\end{remark}

\subsection{Asymptotic Rates of RLL Sequences}
\label{sec:asymptotics}

The number of RLL sequences, denoted $  S_{L}(n)  =  | \myS_{L}(n) | $,
obeys the recurrence relation
$ S_{L}(n) = \sum_{\ell \in L}  S_{L}(n - \ell) $ with initial conditions
$ S_{L}(0) = 1 $ and $ S_{L}(n) = 0 $ for $ n < 0 $, which implies that
$ S_{L}(n) \sim c \lambda^{-n} $ as $ n \to \infty $, where $ \lambda $
is the unique positive solution of the characteristic equation
$ \sum_{\ell \in L} x^{\ell} = 1 $ (see, e.g., \cite{immink, tang+bahl}).
Therefore, the number of bits of information such sequences contain equals
\begin{equation}
\label{eq:rllcap}
  \log S_{L}(n) = - n \log\lambda + {\mathcal O}(1) ,
\end{equation}
and their ``capacity''---the exponential growth rate of $ \mySr_L(n) $ in
base $ 2 $---equals $ \lim_{n\to\infty} \frac{1}{n} \log S_{L}(n) = -\log\lambda $
bits per symbol.
We shall refine this statement below by considering RLL sequences with
additional constraints on their Hamming weight and the number of runs
(building blocks) they contain.

\begin{figure}%[h]
 \centering
  \includegraphics[width=\columnwidth]{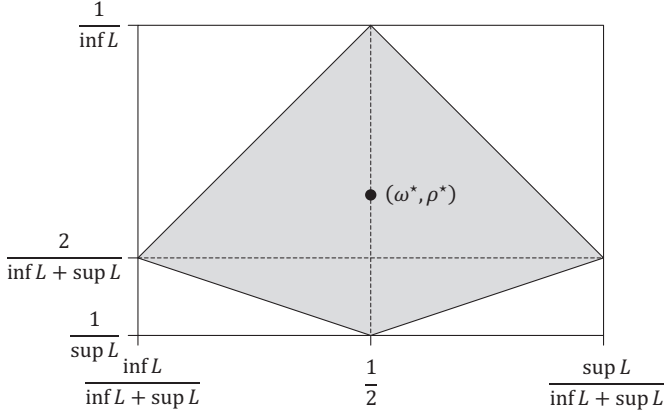}
\caption{The region of the allowed values $ (\omega, \rho) $ -- relative
Hamming weight and relative number of runs -- in binary sequences with
runlengths restricted to a set $ L \subseteq \N $ (see \eqref{eq:omegarhobounds}).
The point $ (\omega^\star, \rho^\star) $ represents the typical values
of these parameters (see Section \ref{sec:typical} ahead).}
\label{fig:region}
\end{figure}

Define
\begin{equation}
\label{eq:rllrw}
  \myS_L(n, w, r)  =  \big\{ \myx \in \myS_L(n) : \weight(\myx) = w , \, \run(\myx) = r \big\}
\end{equation}
and $ S_L(n, w, r) = \big| \myS_L(n, w, r) \big| $.
We may assume that
\begin{align}
\label{eq:nrbounds}
  r \lmin  \leqslant  n  \leqslant  r \lmax
\end{align}
and
\begin{subequations}
\label{eq:wrbounds}
\begin{alignat}{2}
  \left\lfloor r/2 \right\rfloor \lmin  &\leqslant  \phantom{n,} w
	&&\leqslant  \left\lceil r/2 \right\rceil \lmax ,  \\
  \left\lfloor r/2 \right\rfloor \lmin  &\leqslant  n - w
	&&\leqslant  \left\lceil r/2 \right\rceil \lmax ,
\end{alignat}
\end{subequations}
as otherwise $ \myS_{L}(n, w, r) $ is empty.
In the asymptotic regime $ n \to \infty $, $ w \sim \omega n $, $ r \sim \rho n $,
for fixed $ \omega, \rho $, \eqref{eq:nrbounds} and \eqref{eq:wrbounds}
imply that the parameters $ (\omega, \rho) $ are restricted to the region
\begin{subequations}
\label{eq:omegarhobounds}
\begin{align}
\label{eq:wrbounds1}
  \frac{1}{\lmax}  \leqslant\  &\rho  \leqslant  \frac{1}{\lmin} ,  \\
\label{eq:wrbounds2}
	\max\! \Big\{ \frac{\rho}{2} \lmin, \, 1 - \frac{\rho}{2} \lmax \Big\}  \leqslant\  &\omega  \leqslant   \\
\nonumber
	& \min\! \Big\{ \frac{\rho}{2} \lmax, \, 1 - \frac{\rho}{2} \lmin \Big\} .
\end{align}
\end{subequations}
This region, depicted in Figure \ref{fig:region}, can also be represented as
\begin{subequations}
\label{eq:omegarhobounds2}
\begin{align}
\label{eq:wrbounds3}
  \frac{\lmin}{\lmin + \lmax}  \leqslant\  &\omega  \leqslant  \frac{\lmax}{\lmin + \lmax} ,  \\
\label{eq:wrbounds4}
	\frac{2}{\lmax} \max\! \big\{ \omega, \, 1 - \omega \big\}  \leqslant\  &\rho  \leqslant
 \frac{2}{\lmin} \min\! \big\{ \omega, \, 1 - \omega \big\} ,
\end{align}
\end{subequations}
where it is understood that $ \frac{\lmax}{\lmin + \lmax} = 1 $ when $ \lmax = \infty $.

The following statement gives a characterization of the asymptotic rate,
or information capacity, of the RLL sequences from $ \myS_L(n, w, r) $, as
well as the second-order term in the asymptotic expansion of their rate,
for any point $ (\omega, \rho) $ in the interior of the region \eqref{eq:omegarhobounds2}.
For simplicity, here and hereafter, we write $ S_{L}(n, \omega n, \rho n) $
instead of, e.g., $ S_{L}(n, \lfloor \omega n \rfloor, \lfloor \rho n \rfloor) $.

\begin{theorem}%[Asymptotics of $ S_L(n, w, r) $]
\label{thm:rll}
Fix any 
$ \omega \in \big( \frac{\lmin}{\lmin + \lmax}, \frac{\lmax}{\lmin + \lmax} \big) $
and $ \rho \in \big( \frac{2}{\lmax} \max\! \big\{ \omega, 1 - \omega \big\},
      \frac{2}{\lmin} \min\! \big\{ \omega, 1 - \omega \big\} \big) $.
As $ n \to \infty $,
\begin{equation}
\label{eq:rllasymptotics}
  \log S_{L}(n, \omega n, \rho n)
	 =  n \sigma_{L}(\omega, \rho) - \log n + {\mathcal O}(1) ,
\end{equation}
where
\begin{equation}
\label{eq:rllcapacity}
\begin{aligned}
  &\sigma_{L}(\omega, \rho)  \\
  &=  - (1 - \omega) \log \alpha - \omega \log \beta
	    + \frac{\rho}{2} \log\!\left( \sum_{\ell \in L} \alpha^\ell \sum_{\ell' \in L} \beta^{\ell'} \right)
\end{aligned}
\end{equation}
and $ (\alpha, \beta) $ is the unique pair of positive real numbers satisfying
the equations%
\footnote{The dependence of $ \alpha $ and $ \beta $ on $ \omega $, $ \rho $
and $ L $ is not made explicit for reasons of notational simplicity.}
\begin{subequations}
\label{eq:alphabeta}
\begin{align}
  \sum_{\ell \in L} \left( \ell - \frac{2 (1 - \omega)}{\rho} \right) \alpha^\ell  &=  0,  \\
  \sum_{\ell \in L} \left( \ell - \frac{2 \omega}{\rho} \right) \beta^{\ell}  &=  0 .
\end{align}
\end{subequations}
\end{theorem}
\begin{IEEEproof}
The following recurrence relation is valid for all sufficiently large
$ n, w, r $:
\begin{equation}
\label{eq:rllrec}
  S_L(n, w, r)
	 =  \sum_{\ell \in L} \sum_{\ell' \in L}  S_L(n - \ell - \ell', w - \ell', r - 2) .
\end{equation}
This follows from the fact that every sequence in $ \myS_{L}(n, w, r) $
can be obtained by appending the blocks $ 0^\ell $ and $ 1^{\ell'} $ to
a sequence in $ \myS_{L}(n - \ell - \ell', w - \ell', r - 2) $, and that
the order in which the two blocks are appended is uniquely determined by
the last symbol of the latter sequence.
The statement we are after can now be derived from the characteristic
equation of the relation \eqref{eq:rllrec}, namely
\begin{equation}
\label{eq:G}
  G(x, y, z)  =  1 - \sum_{\ell \in L} \sum_{\ell' \in L} x^{\ell + \ell'} y^{\ell'} z^2  =  0 ,
\end{equation}
by using the methods of multivariate analytic combinatorics.
In particular, \cite[Theorem 1.3]{pemantle+wilson1} implies that, in
the asymptotic regime under consideration,
\begin{align}
\label{eq:rllasymptotics2}
    \log S_{L}(n, w, r)
	 =  \log\!\bigg( \frac{ \tilde{\alpha}^{-n} \tilde{\beta}^{-w} \tilde{\gamma}^{-r} }{ n } \bigg) + {\mathcal O}(1) ,
\end{align}
where $ \big( \tilde{\alpha}, \tilde{\beta}, \tilde{\gamma} \big) $ is the
unique triple of positive real numbers satisfying the system of equations:
$ G(x, y, z) = 0 $ (see \eqref{eq:G}),
$ r x \frac{\partial G}{\partial x} = n z \frac{\partial G}{\partial z} $,
and $ r y \frac{\partial G}{\partial y} = w z \frac{\partial G}{\partial z} $,
i.e., the positive numbers $ \tilde{\alpha}, \tilde{\beta}, \tilde{\gamma} $
are uniquely determined by
\begin{subequations}
\label{eq:abc}
\begin{align}
  \sum_{\ell \in L} \sum_{\ell' \in L}
	\tilde{\alpha}^{\ell + \ell'} \tilde{\beta}^{\ell'} \tilde{\gamma}^2  &=  1  \\
\label{eq:p1}
	r \sum_{\ell \in L} \sum_{\ell' \in L}
	(\ell + \ell') \tilde{\alpha}^{\ell + \ell'} \tilde{\beta}^{\ell'}  &=
	2 n \sum_{\ell \in L} \sum_{\ell' \in L}
	\tilde{\alpha}^{\ell + \ell'} \tilde{\beta}^{\ell'}  \\
\label{eq:p2}
	r \sum_{\ell \in L} \sum_{\ell' \in L}
	\ell' \tilde{\alpha}^{\ell + \ell'} \tilde{\beta}^{\ell'}  &=
	2 w \sum_{\ell \in L} \sum_{\ell' \in L}
	\tilde{\alpha}^{\ell + \ell'} \tilde{\beta}^{\ell'}  .
\end{align}
\end{subequations}
As $ \frac{w}{n} \to \omega $, $ \frac{r}{n} \to \rho $, we have
$ \tilde{\alpha} \to \alpha $, $ \tilde{\alpha} \cdot \tilde{\beta} \to \beta $,
$ \tilde{\gamma} \to \gamma $, where $ (\alpha, \beta, \gamma) $ is the
unique triple of positive numbers satisfying
\begin{subequations}
\label{eq:abc2}
\begin{align}
\label{eq:pom1}
  \sum_{\ell \in L} \alpha^{\ell} \sum_{\ell' \in L} \beta^{\ell'}  &=  \gamma^{-2}  \\
\label{eq:pom2}
	\rho \sum_{\ell \in L} \ell \alpha^\ell  &=  2 (1 - \omega) \sum_{\ell \in L} \alpha^\ell  \\
\label{eq:pom3}
	\rho  \sum_{\ell' \in L} \ell' \beta^{\ell'}  &=  2 \omega  \sum_{\ell' \in L} \beta^{\ell'}
\end{align}
\end{subequations}
(\eqref{eq:pom2} is obtained by subtracting \eqref{eq:p2} from \eqref{eq:p1}),
and then \eqref{eq:rllasymptotics2} reduces to
\begin{align}
\label{eq:rllasymptotics3}
    \log S_{L}(n, \omega n, \rho n)
	 =  \log\!\left( \frac{ \alpha^{-(1-\omega)n} \beta^{-\omega n} \gamma^{-\rho n} }{ n } \right) + {\mathcal O}(1) .
\end{align}
Now just note that \eqref{eq:rllasymptotics3} is equivalent to \eqref{eq:rllasymptotics}
(use \eqref{eq:pom1} to express $ \gamma $ in terms of $ \alpha $ and $ \beta $),
and \eqref{eq:pom2}--\eqref{eq:pom3} to \eqref{eq:alphabeta}.
\end{IEEEproof}
\vspace{2mm}

To find the asymptotic rates of the sequences from $ \myS_L(n, w, r) $
at the points $ (\omega, \rho) $ lying on the boundary of the region
\eqref{eq:omegarhobounds2}, we turn this into a one-dimensional problem
wherein one of the parameters $ w, r $ is a function of the other.
For example, for the case
$ \omega \in \big( \frac{\lmin}{\lmin+\lmax}, \frac{1}{2} \big) $,
$ \rho = \frac{2}{\lmin} \omega $ (the upper-left boundary in Figure
\ref{fig:region}), consider the quantity
$ S_L(n, \lfloor r/2 \rfloor \lmin, r) $, which is the number of RLL
sequences in which every run of $ 1 $'s is of fixed length $ \lmin $.
To simplify the derivation, consider only even $ r $, introduce auxiliary
variables $ r' = \frac{r}{2} $ and $ n' = n - \frac{r}{2} \lmin $, and denote
$ S'(n', r') = S_L\big(n' + r' \lmin, r' \lmin, 2r'\big) = S_L\big(n, \frac{r}{2} \lmin, r\big) $.
It is then easy to see that the bivariate sequence $ \big( S'(n', r') \big)_{n',r'} $
satisfies the recurrence relation
\begin{equation}
\label{eq:S1}
  S'(n', r') = \sum_{\ell \in L} S'(n'-\ell, r'-1) ,
\end{equation}
the characteristic equation of which is
\begin{align}
\label{eq:char}
  1  =  \sum_{\ell \in L} x^{\ell} y .
\end{align}
Such a sequence is called a generalized Riordan array \cite[Section 12.2]{pemantle+wilson}.
It follows from \cite[Theorem 12.2.2]{pemantle+wilson} that, as $ n' \to \infty $
and $ r' \to \infty $ with $ \frac{n'}{r'} $ (approximately) constant,
\begin{align}
\label{eq:S'}
  S'(n', r')  \,\sim\,
	 \frac{c}{\sqrt{n'}}  2^{- n' \log \tilde{\alpha} + r' \log v(\tilde{\alpha}) } ,
\end{align}
where $ v(x) = \sum_{\ell \in L} x^\ell $ (see \eqref{eq:char}) and $ \tilde{\alpha} $
is the unique positive solution of the equation $ r' x \frac{\partial v}{\partial x} = n' v(x) $, i.e.,
\begin{equation}
\label{eq:alpha'}
  \sum_{\ell \in L} \left( \ell - \frac{n'}{r'} \right) \tilde{\alpha}^\ell = 0 .
\end{equation}
Now recalling what $ n' $, $ r' $, and $ S'(n', r') $ stand for, we can
find from \eqref{eq:S'} and \eqref{eq:alpha'} the desired asymptotics.
By carrying out a similar analysis for the remaining boundaries of the
region \eqref{eq:omegarhobounds2}, we obtain the following statement.

\begin{theorem}
As $ n \to \infty $, the relation
\begin{align}
\label{eq:rllasympboundary1}
  \log S_{L}\!\left( n, \omega n, \rho n \right)
	 =  n \sigma_{L}\!\left( \omega, \rho \right) - \frac{1}{2} \log n + {\mathcal O}(1)
\end{align}
holds for:
\begin{itemize}
\item
$ \omega \in \big( \frac{\lmin}{\lmin+\lmax}, \frac{1}{2} \big) $,
$ \rho = \frac{2}{\lmin} \omega $, where
\begin{align}
\label{eq:expb1}
	\sigma_{L}\!\left( \omega, \frac{2}{\lmin} \omega \right)
	 = - (1 - \omega) \log \alpha + \frac{\omega}{\lmin} \log \sum_{\ell \in L} \alpha^{\ell}
\end{align}
and $ \alpha $ is the unique positive real number satisfying
\begin{equation}
\label{eq:alphab1}
  \sum_{\ell \in L} \left( \ell - \frac{1 - \omega}{\omega} \lmin \right) \alpha^{\ell} = 0 ;
\end{equation}
\item
$ \omega \in \big( \frac{1}{2} , \frac{\lmax}{\lmin+\lmax} \big) $,
$ \rho = \frac{2}{\lmin} (1 - \omega) $, where
\begin{align}
\label{eq:expb2}
	\sigma_{L}\!\left( \omega, \frac{2}{\lmin} (1 - \omega) \right)
	 = - \omega \log \alpha + \frac{1 - \omega}{\lmin} \log \sum_{\ell \in L} \alpha^{\ell}
\end{align}
and $ \alpha $ is the unique positive real number satisfying
\begin{equation}
\label{eq:alphab2}
  \sum_{\ell \in L} \left( \ell - \frac{\omega}{1 - \omega} \lmin \right) \alpha^{\ell} = 0 ;
\end{equation}
\item
$ \omega \in \big( \frac{\lmin}{\lmin+\lmax}, \frac{1}{2} \big) $,
$ \rho = \frac{2}{\lmax} (1 - \omega) $, $ \lmax < \infty $, where
\begin{align}
\label{eq:expb3}
	\sigma_{L}\!\left( \omega, \frac{2}{\lmax} (1 - \omega) \right)
	 = - \omega \log \alpha + \frac{1 - \omega}{\lmax} \log \sum_{\ell \in L} \alpha^{\ell}
\end{align}
and $ \alpha $ is the unique positive real number satisfying
\begin{equation}
\label{eq:alphab3}
  \sum_{\ell \in L} \left( \ell - \frac{\omega}{1 - \omega} \lmax \right) \alpha^{\ell} = 0 ;
\end{equation}
\item
$ \omega \in \big( \frac{1}{2} , \frac{\lmax}{\lmin+\lmax} \big) $,
$ \rho = \frac{2}{\lmax} \omega $, $ \lmax < \infty $, where
\begin{align}
\label{eq:expb4}
	\sigma_{L}\!\left( \omega, \frac{2}{\lmax} \omega \right)
	 = - (1 - \omega) \log \alpha + \frac{\omega}{\lmax} \log \sum_{\ell \in L} \alpha^{\ell}
\end{align}
and $ \alpha $ is the unique positive real number satisfying
\begin{equation}
\label{eq:alphab4}
  \sum_{\ell \in L} \left( \ell - \frac{1 - \omega}{\omega} \lmax \right) \alpha^{\ell} = 0 .
\end{equation}
\end{itemize}
\end{theorem}

In the last two cases, if $ \lmax = \infty $ the two boundaries
degenerate into one ($ \rho = 0 $), in which case we define
$ \sigma_{L}(\omega, 0) = 0 $ for all
$ \omega \in \big[ \frac{\lmin}{\lmin + \lmax}, \frac{\lmax}{\lmin + \lmax} \big] $.
At the corner-points of the region \eqref{eq:omegarhobounds2}, we set
$ \sigma_{L}\big(\frac{\lmin}{\lmin + \lmax}, \frac{2}{\lmin + \lmax}\big) =
	\sigma_{L}\big(\frac{\lmax}{\lmin + \lmax}, \frac{2}{\lmin + \lmax}\big) = 
	\sigma_{L}\big(\frac{1}{2}, \frac{1}{\lmin}\big) =
  \sigma_{L}\big(\frac{1}{2}, \frac{1}{\lmax}\big) = 0 $
by continuous extension.
Finally, when it is necessary to assign value to $ \sigma_{L}(\omega, \rho) $
for $ (\omega, \rho) $ outside the region \eqref{eq:omegarhobounds2}, we may
write $ \sigma_{L}(\omega, \rho) = - \infty $.

\begin{example}[Unconstrained case]
\textnormal{
  For $ L = \N $ it is possible obtain an explicit expression for the capacity.
By using the identities $ \sum_{\ell=1}^\infty x^\ell = \frac{x}{1-x} $,
$ \sum_{\ell=1}^\infty \ell x^\ell = \frac{x}{(1-x)^2} $, one
can solve \eqref{eq:alphabeta} to get
$ \alpha = 1 - \frac{ \rho }{ 2 (1 - \omega) } $,
$ \beta = 1 - \frac{ \rho }{ 2 \omega } $, and hence
\begin{equation}
\label{eq:sigmaN}
  \sigma_{\N}(\omega, \rho)
  = (1 - \omega) H\!\left( \frac{ \rho }{ 2 (1 - \omega) } \right)
	  + \omega H\!\left( \frac{ \rho }{ 2 \omega } \right) ,
\end{equation}
for any $ \omega \in [0, 1] $, $ \rho \in [0, 2 \min\{\omega, 1 - \omega\}] $.
}
\myqed
\end{example}

\subsection{Weight Only and Run Only Constraints}
\label{sec:wr}

In this subsection we derive the asymptotic rates of RLL sequences when
only one of the parameters $ w, r $ is restricted.
These special cases are arguably more likely to be of relevance in applications.

\subsubsection{Constant-Weight RLL Sequences}

Define
\begin{equation}
\label{eq:rllw}
\begin{aligned}
	\myS_L(n, w, \star)
	 &=  \bigcup_{r} \myS_L(n, w, r) \\
	 &= \big\{ \myx \in \myS_L(n) : \weight(\myx) = w \big\} ,
\end{aligned}
\end{equation}
and $ S_L(n, w, \star) = \big| \myS_L(n, w, \star) \big| $.
Starting from the relation
\begin{equation}
\label{eq:recrllw}
  S_{L}(n, w, \star)
	 = \sum_{\ell \in L} \sum_{\ell' \in L} S_{L}(n - \ell - \ell', w - \ell', \star) ,
\end{equation}
and applying \cite[Theorem 1.3]{pemantle+wilson1}, as we did in the proof
of Theorem \ref{thm:rll}, we obtain the following statement.

\begin{theorem}
\label{thm:rllw}
Fix $ \omega \in \big( \frac{\inf L}{\inf L + \sup L}, \frac{\sup L}{\inf L + \sup L} \big) $.
As $ n \to \infty $,
\begin{equation}
\label{eq:expw}
  \log S_{L}(n, \omega n, \star)
	 =  n \sigma_{L}(\omega, \star) - \frac{1}{2} \log n + {\mathcal O}(1) ,
\end{equation}
where
\begin{equation}
\label{eq:expW2}
  \sigma_{L}(\omega, \star)  =  - (1 - \omega) \log \alpha - \omega \log \beta ,
\end{equation}
and $ (\alpha, \beta) $ is the unique pair of positive real numbers satisfying
the system of equations:
\begin{subequations}
\label{eq:alphabetaw}
\begin{align}
  \sum_{\ell \in L} \alpha^\ell \sum_{\ell' \in L} \beta^{\ell'}  
	&=  1  \\
  \omega \sum_{\ell \in L} \ell \alpha^\ell \sum_{\ell' \in L} \beta^{\ell'}  
	&=  (1 - \omega) \sum_{\ell \in L} \alpha^\ell \sum_{\ell' \in L} \ell' \beta^{\ell'}  .
\end{align}
\end{subequations}
\end{theorem}

We also define
$ \sigma_{L}\big( \frac{\inf L}{\inf L + \sup L}, \star \big)
  = \sigma_{L}\big( \frac{\sup L}{\inf L + \sup L}, \star \big) = 0 $
by continuous extension.

\begin{example}[Unconstrained case, continued]
\textnormal{
  For $ L = \N $, the solution to \eqref{eq:alphabetaw}
is $ \alpha = 1 - \omega $, $ \beta = \omega $, and hence
$ \sigma_{\N}(\omega, \star) = H(\omega) $, as expected.
}
\myqed
\end{example}

It follows from the relation $ S_L(n, w, \star) = \sum_{r} S_L(n, w, r) $
and the pigeon-hole principle, that the exponent $ \sigma_{L}(\omega, \star) $,
which represents the information capacity of constant-weight RLL sequences,
can also be obtained as
\begin{align}
\label{eq:wviawr}
  \sigma_{L}(\omega, \star)  =  \sup_{\rho}  \sigma_{L}(\omega, \rho) .
\end{align}
The capacity of RLL sequences with no weight constraints can then be
obtained as
\begin{equation}
  \mysr_{L}(\star, \star)
	 = \sup_{\omega} \mysr_{L}(\omega, \star)
	 = \sup_{\omega, \rho} \mysr_{L}(\omega, \rho) ,
\end{equation}
which, as we already know (see \eqref{eq:rllcap}), equals
\begin{equation}
\label{eq:sigmamax}
  \mysr_{L}(\star, \star)  =  -\log \lambda ,
\end{equation}
where $ \lambda $ is the unique positive number satisfying
\begin{equation}
\label{eq:lambda}
  \sum_{\ell \in L} \lambda^\ell = 1 .
\end{equation}

\subsubsection{Constant-Number-of-Runs RLL Sequences}
\label{sec:ronly}

Define
\begin{equation}
\label{eq:rllr}
\begin{aligned}
  \myS_L(n, \star, r)
	 &=  \bigcup_{w} \myS_L(n, w, r)  \\
	 &= \big\{ \myx \in \myS_L(n) : \run(\myx) = r \big\} ,
\end{aligned}
\end{equation}
and $ S_L(n, \star, r) = \big| \myS_L(n, \star, r) \big| $.
$ \myS_{L}(n, \star, r) $ is the set of all sequences of length $ n $
that are formed by concatenating exactly $ r $ blocks from
$ \{ 0^\ell : \ell \in L \} $ and $ \{ 1^\ell : \ell \in L \} $ in an
alternating manner.
Due to \eqref{eq:dk}, studying the asymptotic behavior of the sequences
from $ \myS_{L}(n, \star, r) $ is essentially equivalent to studying the
asymptotic behavior of \emph{constant-weight sequences with constrained
runs of zeros}.

It is seen from the definition of $ \mySSr_{L}(n, \star, r) $ that the
bivariate sequence $ \big( \mySr_{L}(n, \star, r) \big)_{n,r} $ obeys
the recurrence relation
\begin{equation}
\label{eq:recrllr}
  \mySr_{L}(n, \star, r)  =  \sum_{\ell \in L}  \mySr_{L}(n - \ell, \star, r - 1) .
\end{equation}
from which the following statement can be obtained in the same way as in
\eqref{eq:S1}--\eqref{eq:alpha'}.

\begin{theorem}
\label{thm:asymp}
Fix $ \rho \in \big( \frac{1}{\lmax} , \frac{1}{\lmin} \big) $.
As $ n \to \infty $,
\begin{align}
\label{eq:Sexp}
  \log \mySr_{L}(n, \star, \rho n)
	 =  n\, \mysr_{L}(\star, \rho)  -  \frac{1}{2} \log n  +  {\mathcal O}(1) ,
\end{align}
where
\begin{align}
\label{eq:expr}
	\sigma_{L}(\star, \rho)  =  - \log \alpha + \rho \log \sum_{\ell \in L} \alpha^{\ell}
\end{align}
and $ \alpha $ is the unique positive real number satisfying
\begin{equation}
\label{eq:rho2}
  \sum_{\ell \in L} \left( \ell - \frac{1}{\rho} \right) \alpha^{\ell} = 0 .
\end{equation}
\end{theorem}

We also define
$ \sigma_{L}\!\big( \star, \frac{1}{\lmax} \big)
  = \sigma_{L}\!\big( \star, \frac{1}{\lmin} \big) = 0 $.
The quantity $ \mysr_{L}(\star, \rho) $ just introduced represents the
information capacity of constant-weight binary sequences in which successive
ones are required to be separated by runs of zeros whose lengths are constrained
to the set $ L - 1 $ (see Remark~\ref{rem:dk}).
This exponent was determined in \cite{kovacevic2} for the special case
of $ (d, k) $-sequences, i.e., for $ L = \{ d + 1, \ldots, k + 1 \} $.

\begin{example}[Unconstrained case, continued]
\textnormal{
  For $ L = \N $, the solution to \eqref{eq:rho2} is $ \alpha = 1 - \rho $,
and hence $ \sigma_{\N}(\star, \rho) = H(\rho) $, as expected.
}
\myqed
\end{example}

Analogously to \eqref{eq:wviawr}, $ \sigma_{L}(\star, \rho) $ can be
expressed as
\begin{align}
  \sigma_{L}(\star, \rho)  =  \sup_{\omega}  \sigma_{L}(\omega, \rho) .
\end{align}
As an aside, we note that it is also possible to express the exponent
$ \sigma_{L}(\omega, \rho) $ in terms of $ \sigma_{L}(\star, \rho) $.
To show this, first note that $ S_{L}(n, \star, r) = 2 C_L(n, r) $, where
$ C_L(n, r) $ is the number of $ r $-part \emph{compositions}%
\footnote{An $ m $-part composition of the number $ n $, with parts
restricted to the set $ L \subseteq \N $, is an $ m $-tuple from $ L^m $
summing to $ n $ \cite{heubach+mansour}.}
of the number $ n $, where the parts are restricted to the set $ L $
(the factor $ 2 $ comes from the fact that, given the lengths of the
constituent blocks, the initial symbol is left to be specified in order
to identify the sequence uniquely).
Further note that the quantity $ S_L(n, w, r) $ can also be expressed in
terms of integer compositions as follows:
\begin{equation}
\label{eq:comp}
\begin{aligned}
  S_L(n, w, r) =\, &C_L(n-w, \lfloor r/2 \rfloor) \cdot C_L(w, \lceil r/2 \rceil)  \\
 	                 &+ C_L(n-w, \lceil r/2 \rceil) \cdot C_L(w, \lfloor r/2 \rfloor) .
\end{aligned}
\end{equation}
Namely, if an RLL sequence starts with a $ 1 $ and is built from $ r $
blocks, then it contains $ \lceil r/2 \rceil $ blocks of $ 1 $'s and
$ \lfloor r/2 \rfloor $ blocks of $ 0 $'s.
The lengths of the blocks of $ 1 $'s sum to $ w $, and those of $ 0 $'s
sum to $ n - w $.
This gives the first summand in \eqref{eq:comp}.
Similarly, the second summand counts the RLL sequences that start
with a $ 0 $.
We conclude that
\begin{equation}
\label{eq:Swrviacomp}
\begin{aligned}
  S_L(n, w, r) = \frac{1}{4} \Big( & S_L(n-w, \star, \lfloor r/2 \rfloor) \cdot S_L(w, \star, \lceil r/2 \rceil)  \\
 	                 &+ S_L(n-w, \star, \lceil r/2 \rceil) \cdot S_L(w, \star, \lfloor r/2 \rfloor) \Big) ,
\end{aligned}
\end{equation}
and therefore
\begin{equation}
\label{eq:rllruns}
  \sigma_{L}(\omega, \rho)
	 =  (1 - \omega) \sigma_{L}\!\left( \star, \frac{\rho}{2(1 - \omega)} \right)  +
			\omega \sigma_{L}\!\left( \star, \frac{\rho}{2\omega} \right) ,
\end{equation}
for every $ (\omega, \rho) $ in the region \eqref{eq:omegarhobounds2}.

\subsection{Probabilistic Characterization of the Capacity}
\label{sec:probabilistic}

The capacity function $ \sigma_{L}(\omega, \rho) $ can also be described
by using probabilistic terminology customary in information theory.
It is known that finding the asymptotic rate of RLL sequences is equivalent
to maximizing the \emph{average number of bits (i.e., entropy) per symbol}
at the output of an i.i.d.\ source with alphabet $ L $ \cite[Theorem 1]{zehavi+wolf},
that is,
\begin{equation}
\label{eq:maxent}
  \sigma_L(\star, \star) = \sup_{P_{\mathsf{L}}} \frac{ H(\mathsf{L}) }{ \mathbb{E}[\mathsf{L}] } ,
\end{equation}
where $ \mathsf{L} $ denotes a generic random variable taking values in $ L $,
and $ P_{\mathsf{L}} $ denotes its distribution.
It is also known \cite{zehavi+wolf} that the optimal distribution in
\eqref{eq:maxent} is $ P^\star_{\mathsf{L}}(\ell) = \lambda^\ell $, $ \ell \in L $,
where $ \lambda $ is defined in \eqref{eq:lambda}.
Since we are imposing additional constraints on the blocks of $ 1 $'s,
and therefore also on the blocks of $ 0 $'s, through the Hamming weight,
we shall need a slightly different way of writing \eqref{eq:maxent}, namely
$ \sigma_L(\star, \star) = \sup_{P_{\mathsf{L}_0, \mathsf{L}_1}}
  \frac{ H(\mathsf{L}_0, \mathsf{L}_1) }{ \mathbb{E}[\mathsf{L}_0 + \mathsf{L}_1] } $,
where $ \mathsf{L}_0, \mathsf{L}_1 $ are independent random variables
taking values in $ L $.
The following proposition refines this statement and gives the corresponding
characterization for sequences with fixed weight and number of runs.

\begin{theorem}%[Probabilistic characterization of the capacity]
\label{thm:maxentrll}
Let $ \mathsf{L}_0 $, $ \mathsf{L}_1 $ denote two generic independent
random variables taking values in $ L $, and let $ P_{\mathsf{L}_0} $,
$ P_{\mathsf{L}_1} $denote their distributions.
For every $ (\omega, \rho) $ in the region \eqref{eq:omegarhobounds2},
\begin{subequations}
\label{eq:maxentropy}
\begin{align}
\nonumber
  &\sigma_{L}(\omega, \rho)  \\
\label{eq:maxentropy1}
	 &= \sup_{ P_{\mathsf{L}_{0}}, P_{\mathsf{L}_{1}} :\,
	          \mathbb{E}[\mathsf{L}_0] = \frac{2 (1-\omega)}{\rho} ,\,
						\mathbb{E}[\mathsf{L}_1] = \frac{2 \omega}{\rho} }
	   \frac{ H(\mathsf{L}_{0}, \mathsf{L}_{1}) }{ \mathbb{E}[\mathsf{L}_{0} + \mathsf{L}_{1}] }  \\
\label{eq:maxentropy2}
	 &= \frac{\rho}{2} \left(
	      \sup_{ P_{\mathsf{L}_{0}}:\, \mathbb{E}[\mathsf{L}_0] = \frac{2 (1-\omega)}{\rho} }  H(\mathsf{L}_{0})  +
				\sup_{ P_{\mathsf{L}_{1}}:\, \mathbb{E}[\mathsf{L}_1] = \frac{2 \omega}{\rho} }  H(\mathsf{L}_{1}) \right) .
\end{align}
\end{subequations}
\end{theorem}
\begin{IEEEproof}
That the expressions in \eqref{eq:maxentropy1} and \eqref{eq:maxentropy2} are
equal follows from the fact that $ \mathsf{L}_0 $ and $ \mathsf{L}_1 $ are independent
and so $ H(\mathsf{L}_0, \mathsf{L}_1) = H(\mathsf{L}_0) + H(\mathsf{L}_1) $,
and that $ \mathbb{E}[\mathsf{L}_0] + \mathbb{E}[\mathsf{L}_1] = \frac{2}{\rho} $
by the optimization constraint.

Consider a point $ (\omega, \rho) $ in the interior of the region \eqref{eq:omegarhobounds2}.
The optimal distributions in \eqref{eq:maxentropy2} are by standard methods
(e.g., Lagrange multipliers, see \cite[Section 12.1]{cover}) found to be of
the form $ P^\star_{\mathsf{L}_0}(\ell) = c_0 \alpha^\ell $,
$ P^\star_{\mathsf{L}_1}(\ell) = c_1 \beta^\ell $, where
$ c_0 = \big( \sum_{\ell \in L} \alpha^\ell \big)^{-1} $,
$ c_1 = \big( \sum_{\ell \in L} \beta^\ell \big)^{-1} $ are the normalizing
constants, and $ \alpha $, $ \beta $ are determined by the conditions
\begin{subequations}
\begin{align}
  \mathbb{E}[\mathsf{L}_0] &= c_0 \sum_{\ell \in L} \ell \alpha^\ell = \frac{2(1 - \omega)}{\rho} ,  \\
	\mathbb{E}[\mathsf{L}_1] &= c_1 \sum_{\ell \in L} \ell \beta^\ell  = \frac{2\omega}{\rho} .
\end{align}
\end{subequations}
After writing out the entropy of each of these distributions explicitly,
one finds that the expression in \eqref{eq:maxentropy2} is equal to that
in \eqref{eq:rllcapacity}.

For the points $ (\omega, \rho) $ at the boundary of the region \eqref{eq:omegarhobounds2}
the analysis is similar, except the optimal distribution in one (or both)
of the suprema in \eqref{eq:maxentropy2} will be a degenerate distribution
of zero entropy.
For example, for $ \rho = \frac{2}{\lmin} \omega $, there is only one distribution
on $ L $ consistent with the requirement $ \mathbb{E}[\mathsf{L}_1] = 2\omega\rho^{-1} = \lmin $ ---
the one with $ P^\star_{\mathsf{L}_1}(\lmin) = 1 $, $ P^\star_{\mathsf{L}_1}(\ell) = 0 $
for $ \ell \neq \lmin $.
Since $ H(P^\star_{\mathsf{L}_1}) = 0 $, the second summand in \eqref{eq:maxentropy2}
is equal to zero, and again by applying standard optimization methods to the first
summand, the expression \eqref{eq:maxentropy2} is shown to be equal to that in
\eqref{eq:expb1}.
The remaining cases are analyzed in a similar way.
\end{IEEEproof}
\vspace{2mm}

Informally, the expression in \eqref{eq:maxentropy} arises as follows.
Think of random RLL sequences obtained by alternately drawing blocks
of zeros and blocks of ones, randomly and independently of each other
(as there is no inherent dependence between blocks of zeros and ones).
For simplicity, assume that the sequences start with a block of zeros
(this convention does not affect the asymptotic rate).
The information capacity of these sequences, i.e., the number of
\emph{bits per symbol} they contain, equals
$ \frac{ H(\mathsf{L}_{0}, \mathsf{L}_{1}) }{ \mathbb{E}[\mathsf{L}_{0} + \mathsf{L}_{1}] } 
  = \frac{ H(\mathsf{L}_{0}) + H(\mathsf{L}_{1}) }{ \mathbb{E}[\mathsf{L}_{0}] + \mathbb{E}[\mathsf{L}_{1}] } $,
where $ \mathsf{L}_{0}, \mathsf{L}_{1} $ are the random lengths of the
corresponding blocks
(to see this, recall that a block of zeros is necessarily followed by a
block of ones, so for the purpose of determining the capacity one may
think of pairs $ \left( 0^{\mathsf{L}_{0}}, 1^{\mathsf{L}_{1}} \right) $
as elementary building blocks).
We are free to choose the probability distributions
$ P_{\mathsf{L}_{0}}, P_{\mathsf{L}_{1}} $ so as to maximize
this expression, but these distributions must satisfy the stated
constraints.
Namely, since the total number of blocks (runs) in RLL sequences is
required to scale as $ \sim\!\rho n $, the number of blocks of zeros
and the number of blocks of ones both scale as $ \sim\!\frac{\rho}{2} n $.
And since the total length of all the blocks of ones, i.e., the Hamming
weight, is required to scale as $ \sim\!\omega n $, the average length
of blocks of ones should be $ \frac{\omega}{\rho/2} $.
Similarly, the average length of blocks of zeros should be
$ \frac{1 - \omega}{\rho/2} $.

The special cases when only the weight or the number of runs are constrained
are easily obtained from Theorem \ref{thm:maxentrll}.

\begin{corollary}
\label{thm:maxentrllw}
Let $ \mathsf{L}_0 $, $ \mathsf{L}_1 $ denote two generic independent
random variables taking values in $ L $, and let $ P_{\mathsf{L}_0} $,
$ P_{\mathsf{L}_1} $denote their distributions.
For every
$ \omega \in \big[ \frac{\inf L}{\inf L + \sup L}, \frac{\sup L}{\inf L + \sup L} \big] $,
\begin{equation}
\label{eq:maxentropycw}
  \sigma_{L}(\omega, \star)
	 = \sup_{ P_{\mathsf{L}_{0}}, P_{\mathsf{L}_{1}} :\,
	          \mathbb{E}[\mathsf{L}_1] = \omega \mathbb{E}[\mathsf{L}_0+ \mathsf{L}_1] }
	   \frac{ H(\mathsf{L}_{0}, \mathsf{L}_{1}) }{ \mathbb{E}[\mathsf{L}_{0} + \mathsf{L}_{1}] } .
\end{equation}
\end{corollary}

For
$ \omega \in \big( \frac{\inf L}{\inf L + \sup L}, \frac{\sup L}{\inf L + \sup L} \big) $,
the optimal distributions in \eqref{eq:maxentropycw} are given by
$ P^\star_{\mathsf{L}_{0}}(\ell) = \big( \sum_{\ell \in L} \alpha^\ell \big)^{-1} \alpha^\ell $
and
$ P^\star_{\mathsf{L}_{1}}(\ell) = \big( \sum_{\ell \in L} \beta^\ell \big)^{-1} \beta^\ell $,
$ \ell \in L $, where $ \alpha, \beta $ are determined by \eqref{eq:alphabetaw}.

\begin{corollary}
\label{thm:maxent}
Let $ \mathsf{L} $ denote a generic random variable taking values in $ L $,
and let $ P_{\mathsf{L}} $ denote its distribution.
For every $ \rho \in \big[ \frac{1}{\lmax}, \frac{1}{\lmin} \big] $,
\begin{equation}
\label{eq:maxentropycnr}
  \mysr_{L}(\star, \rho)
	 = \sup_{ P_{\mathsf{L}} :\, \mathbb{E}[\mathsf{L}] = \rho^{-1} }
	   \frac{ H(\mathsf{L}) }{ \mathbb{E}[\mathsf{L}] }
	 = \rho \cdot \sup_{ P_{\mathsf{L}} :\, \mathbb{E}[\mathsf{L}] = \rho^{-1} }  H(\mathsf{L}) .
\end{equation}
\end{corollary}

Hence, the requirement that the number of runs scales as $ \sim\!\rho n $
translates, in the probabilistic interpretation, into the requirement that
the average length of the runs be $ 1/\rho $.
For $ \rho \in \big( \frac{1}{\lmax}, \frac{1}{\lmin} \big) $, the
optimal distribution in \eqref{eq:maxentropycnr} is of the form
$ P^\star_{\mathsf{L}}(\ell) = \big( \sum_{\ell \in L} \alpha^\ell \big)^{-1} \alpha^\ell $,
$ \ell \in L $, where $ \alpha $ is determined by \eqref{eq:rho2}.

\subsection{Properties of the Capacity}
\label{sec:properties}

We next state some useful properties of the capacity as a function of its parameters.

\begin{proposition}
\label{thm:properties}
%\phantom{a}
\begin{itemize}
\item[\textnormal{(a)}]
For every fixed $ \rho \in \big[ \frac{1}{\lmax}, \frac{1}{\lmin} \big] $,
the mapping $ \omega \mapsto \mysr_{L}(\omega, \rho) $ is symmetric, continuous,
and strictly concave on the interval \eqref{eq:wrbounds2}.
It attains its maximum value at $ \omega^\star = 1/2 $, and this maximum value
equals $ \mysr_{L}(1/2, \rho) = \mysr_{L}(\star, \rho) $ (see \eqref{eq:expr}).
\item[\textnormal{(b)}]
For every fixed $ \omega \in \big[ \frac{\lmin}{\lmin+\lmax}, \frac{\lmax}{\lmin+\lmax} \big] $,
the mapping $ \rho \mapsto \mysr_{L}(\omega, \rho) $ is continuous and strictly
concave on the interval \eqref{eq:wrbounds4}.
It attains its maximum value at
\begin{equation}
\label{eq:rwstar}
  \rho_\omega^\star
	 = 2 \left( \frac{ \sum_{\ell \in L} \ell \alpha^{\ell} }{ \sum_{\ell \in L} \alpha^{\ell} }
	       + \frac{ \sum_{\ell' \in L} \ell' \beta^{\ell'} }{ \sum_{\ell' \in L} \beta^{\ell'} } \right)^{\!-1} , 
\end{equation}
where $ (\alpha, \beta) $ is the unique pair of positive real numbers
solving the equations
\begin{subequations}
\label{eq:ab2}
\begin{align}
\label{eq:pom11}
  \sum_{\ell \in L} \alpha^{\ell} \sum_{\ell' \in L} \beta^{\ell'}  &=  1  \\
\label{eq:pom12}
	\omega \sum_{\ell \in L} \ell \alpha^\ell  \sum_{\ell' \in L} \beta^{\ell'}
	&=  (1 - \omega) \sum_{\ell \in L} \alpha^\ell  \sum_{\ell' \in L} \ell' \beta^{\ell'} ,
\end{align}
\end{subequations}
and this maximum value equals $ \mysr_{L}(\omega, \rho^\star_\omega) = \mysr_{L}(\omega, \star) $
(see \eqref{eq:expW2}).
\item[\textnormal{(c)}]
The mapping $ \omega \mapsto \mysr_{L}(\omega, \star) $ is symmetric, continuous,
and strictly concave on the interval
$ \omega \in \big[ \frac{\lmin}{\lmin+\lmax}, \frac{\lmax}{\lmin+\lmax} \big] $.
It attains its maximum value at $ \omega^\star = 1/2 $, and this maximum value
equals $ \mysr_{L}(1/2, \star) = \mysr_{L}(\star, \star) $ (see \eqref{eq:sigmamax}).
\item[\textnormal{(d)}]
The mapping $ \rho \mapsto \mysr_{L}(\star, \rho) $ is continuous and strictly
concave on the interval $ \rho \in \big[ \frac{1}{\lmax}, \frac{1}{\lmin} \big] $.
It attains its maximum value at
\begin{equation}
\label{eq:rstar}
  \rho^\star
	 = \rho^\star_{1/2}
	 = \left( \sum_{\ell \in L} \ell \lambda^\ell \right)^{\!-1} , 
\end{equation}
where $ \lambda $ is the unique positive number satisfying
$ \sum_{\ell \in L} \lambda^\ell = 1 $,
and this maximum value equals
$ \mysr_{L}(\star, \rho^\star) = \mysr_{L}(\star, \star) $ (see \eqref{eq:sigmamax}).
\end{itemize}
\end{proposition}
\begin{IEEEproof}
(a) Since $ S_L(n, w, r) = S_L(n, n - w, r) $ (this is seen by replacing
$ 0 $'s with $ 1 $'s, and vice versa, in all sequences),
the function $ \sigma_L(\omega, \rho) $ is symmetric in the argument $ \omega $,
meaning that $ \sigma_L(\omega, \rho) = \sigma_L(1 - \omega, \rho) $.
Further, differentiating \eqref{eq:rllcapacity} with respect to $ \omega $,
and using \eqref{eq:alphabeta}, we get
\begin{align}
\label{eq:derivw}
  \frac{\partial}{\partial \omega} \sigma_{L}(\omega, \rho)
	 &= \log \frac{\alpha}{\beta} ,  \\
  \frac{\partial^2}{\partial \omega^2} \sigma_{L}(\omega, \rho)
	 &= \frac{1}{\ln 2} \left( \frac{1}{\alpha} \frac{\partial \alpha}{\partial \omega}
	    - \frac{1}{\beta} \frac{\partial \beta}{\partial \omega} \right) .
\end{align}
That the second derivative is negative, and hence that $ \sigma_L(\omega, \rho) $
is concave in $ \omega $, is shown by differentiating \eqref{eq:alphabeta},
upon which one obtains
\begin{subequations}
\label{eq:derivabw}
\begin{alignat}{2}
  \frac{1}{\alpha} \frac{ \partial \alpha }{ \partial \omega }
	 &= \frac{ -\frac{2}{\rho} \sum_{\ell \in L} \alpha^\ell }{ \sum_{\ell \in L} \left(\ell - \frac{2(1 - \omega)}{\rho} \right)^{\!2} \alpha^\ell  }  &&<  0 ,  \\
  \frac{1}{\beta} \frac{ \partial \beta }{ \partial \omega }
	 &= \frac{ \frac{2}{\rho} \sum_{\ell \in L} \beta^\ell }{ \sum_{\ell \in L} \left(\ell - \frac{2\omega}{\rho} \right)^{\!2} \beta^\ell  }  &&>  0 .
\end{alignat}
\end{subequations}
Symmetry and concavity together imply that the maximizer of
$ \omega \mapsto \mysr_{L}(\omega, \rho) $ is $ \omega^\star = 1/2 $.

(b) Differentiating \eqref{eq:rllcapacity} with respect to $ \rho $, we get
\begin{align}
\label{eq:derivr}
  \frac{\partial}{\partial \rho} \sigma_{L}(\omega, \rho)
	 &= \frac{1}{2} \log\!\left( \sum_{\ell \in L} \alpha^{\ell} \sum_{\ell' \in L} \beta^{\ell'} \right) ,  \\
\label{eq:derivr2}
  \frac{\partial^2}{\partial \rho^2} \sigma_{L}(\omega, \rho)
	 &= \frac{1}{\ln 2} \left( \frac{1 - \omega}{\rho \alpha} \frac{\partial \alpha}{\partial \rho}
	    + \frac{\omega}{\rho \beta} \frac{\partial \beta}{\partial \rho} \right) .
\end{align}
Equating the first derivative to zero and using \eqref{eq:alphabeta},
we find that the maximizer $ \rho_\omega^\star $ is defined by the set
of equations \eqref{eq:rwstar} and \eqref{eq:ab2}.
That the expression in \eqref{eq:derivr2} is negative, and hence that
$ \sigma_L(\omega, \rho) $ is concave in $ \rho $, is shown by differentiating
\eqref{eq:alphabeta}, upon which one obtains
\begin{subequations}
\label{eq:derivabr}
\begin{alignat}{2}
  \frac{1}{\alpha} \frac{ \partial \alpha }{ \partial \rho }
	 &= \frac{ - \frac{1}{\rho} \sum_{\ell \in L} \ell \alpha^\ell }{ \sum_{\ell \in L} \left(\ell - \frac{2(1 - \omega)}{\rho} \right)^{\!2} \alpha^\ell  }  &&<  0 ,  \\
  \frac{1}{\beta} \frac{ \partial \beta }{ \partial \rho }
	 &= \frac{ - \frac{1}{\rho} \sum_{\ell \in L} \ell \beta^\ell }{ \sum_{\ell \in L} \left(\ell - \frac{2\omega}{\rho} \right)^{\!2} \beta^\ell  }  &&<  0 .
\end{alignat}
\end{subequations}

(c) Note that this does not immediately follow from (a) because
$ \mysr_{L}(\omega, \star) = \mysr_{L}(\omega, \rho^\star_\omega) $ and
$ \rho^\star_\omega $ is itself a function of $ \omega $, but the proof
is carried out in a similar way by noting that
$ S_L(n, w, \star) = S_L(n, n - w, \star) $, and therefore
$ \mysr_{L}(\omega, \star) = \mysr_{L}(1 - \omega, \star) $, and by
differentiating \eqref{eq:expW2} and \eqref{eq:alphabetaw}.

(d) Since $ \mysr_{L}(\star, \rho) = \mysr_{L}(1/2, \rho) $, this statement
follows directly from (b).
\end{IEEEproof}
\vspace{2mm}

Therefore, the capacity function $ \mysr_{L}(\omega, \rho) $ is \emph{uniquely}
maximized at $ (\omega^\star, \rho^\star) $, and the maximum value is given
by \eqref{eq:sigmamax}.
In other words, sequences of weight $ \frac{1}{2} n $ (which are called
``balanced'') and having $ \rho^\star n $ runs achieve the capacity of
RLL sequences with no weight or run constraints, namely%
\begin{equation}
\label{eq:simwstar}
  \log \mySr_{L}(n, \omega^\star n, \rho^\star n)  \,\sim\,  \log \mySr_{L}(n) .
\end{equation}
However, comparing \eqref{eq:rllcap} with \eqref{eq:rllasymptotics}, we see
that these additional constraints incur a ``penalty'' of $ \sim\!\frac{\log n}{n} $
bits per symbol (or $ \sim\!\frac{\log n}{2n} $ in case only one of them is
active, see \eqref{eq:expw} and \eqref{eq:Sexp}), meaning that the convergence
to capacity is slower.

The following proposition states how the capacity depends on $ L $.

\begin{proposition}
\label{thm:properties2}
\begin{itemize}
\item[\textnormal{(a)}]
The mapping $ L \mapsto \mysr_{L}(\omega, \rho) $ is strictly monotone on
the lattice of subsets of $ \N $, meaning that
$ \mysr_{L_1}(\omega, \rho)  <  \mysr_{L_2}(\omega, \rho) $ for any
$ L_1 \subsetneq L_2 \subseteq \N $ and any $ (\omega, \rho) $
in the region \eqref{eq:omegarhobounds2} defined for the set $ L_2 $.
The same is true for the mappings $ L \mapsto \mysr_{L}(\omega, \star) $,
$ L \mapsto \mysr_{L}(\star, \rho) $, and $ L \mapsto \mysr_{L}(\star, \star) $.
The corresponding limiting values are%
\begin{align}
\label{eq:limitwr}
  \sigma_{\N}(\omega, \rho)
   &= (1 - \omega) H\!\left( \frac{ \rho }{ 2 (1 - \omega) } \right)
	    + \omega H\!\left( \frac{ \rho }{ 2 \omega } \right) ,  \\
	\sigma_{\N}(\omega, \star)
	 &= H(\omega) ,  \\
\label{eq:limitr}
	\sigma_{\N}(\star, \rho)
	 &= H(\rho) ,  \\
	\sigma_{\N}(\star, \star)
	 &= 1 .
\end{align}
\item[\textnormal{(b)}]
For every $ L \subseteq \N $, $ s \in \{- \inf L + 1, -\inf L + 2, \ldots \} $,
$ \rho \in \big[ \frac{1}{s + \sup L}, \frac{1}{s + \inf L} \big] $, and
$ \omega \in \big[ \max\! \big\{ \frac{\rho}{2} s + \frac{\rho}{2} \lmin , \linebreak
                   1 - \frac{\rho}{2} s - \frac{\rho}{2} \lmax \big\} ,
								   \min\! \big\{ \frac{\rho}{2} s + \frac{\rho}{2} \lmax,
									 1 - \frac{\rho}{2} s - \frac{\rho}{2} \lmin \big\} \big] $,%
\begin{align}
\label{eq:sL}
  \mysr_{L+s}(\omega, \rho)
	 &= (1 - s \rho) \mysr_{L}\!\left( \frac{ \omega - s \rho / 2 }{ 1 - s \rho} , \frac{ \rho }{ 1 - s \rho} \right) ,  \\
\label{eq:sL2}
  \mysr_{L+s}(\star, \rho)
	 &= (1 - s \rho) \mysr_{L}\!\left( \star , \frac{ \rho }{ 1 - s \rho} \right) .
\end{align}
\item[\textnormal{(c)}]
For every $ L \subseteq \N $,
$ \sigma_{L+1}(\star, \star) < \sigma_L(\star, \star) $.
In other words, $ \sigma_{L+s}(\star, \star) $ is a monotonically
decreasing function of $ s \in \N $.
The limiting value is $ \lim_{s \to \infty} \sigma_{L+s}(\star, \star) = 0 $.
\end{itemize}
\end{proposition}
\begin{IEEEproof}
(a) Consider first the case $ L \mapsto \sigma_L(\star, \rho) $.
For any $ \ell' \notin L $, we have
$ \mySSr_{L\cup\{\ell'\}}(n, \star, r) \supseteq \mySSr_{L}(n, \star, r) $,
and therefore $ \mysr_{L\cup\{\ell'\}}(\star, \rho) \geqslant \mysr_{L}(\star, \rho) $.
That the latter inequality is strict can be seen by considering \eqref{eq:rho2}
and noting that $ \alpha $ cannot remain the same as $ L $ increases.
The monotonicity of $ L \mapsto \sigma_L(\omega, \rho) $ now follows from,
e.g., \eqref{eq:rllruns}, and it further implies the monotonicity of
$ L \mapsto \mysr_{L}(\omega, \star) $ and $ L \mapsto \mysr_{L}(\star, \star) $.
The limiting value \eqref{eq:limitwr} was found in \eqref{eq:sigmaN}.

(b) Suppose for simplicity that $ r $ is even.
A one-to-one correspondence between $ \mySSr_{L}(n, w, r) $ and
$ \mySSr_{L+s}\big( n+sr, w + s \frac{r}{2}, r \big) $ can be established
by extending every run of every sequence from $ \mySSr_{L}(n, w, r) $ by
$ s $ symbols (here adding $ s $ symbols to an existing run means removing
$ |s| $ symbols from it if $ s $ is negative).
This implies that
$ \mySr_{L}(n, w, r) = \mySr_{L+s}\big( n + s r, w + s \frac{r}{2}, r \big) $,
and hence
\begin{equation}
  \mysr_{L}(\omega, \rho)
	 = (1 + s \rho) \mysr_{L+s}\!\left( \frac{\omega + s \rho / 2}{1 + s \rho}, \frac{\rho}{1 + s \rho} \right) .
\end{equation}
This is equivalent to \eqref{eq:sL}.
By maximizing \eqref{eq:sL} over $ \omega $, we get \eqref{eq:sL2}.

(c) Intuitively the statement is clear because increasing the lengths of
the allowed blocks reduces the number of length-$ n $ sequences that can
be built from those blocks.
Formally, we can write $ \sum_{\ell \in L+s} x^\ell = x^s \sum_{\ell \in L} x^\ell $
and conclude from \eqref{eq:lambda} that $ \lambda $ is a monotonically increasing
(and hence $ \mysr_{L}(\star, \star) = -\log \lambda $ monotonically decreasing)
function of the shift $ s $.
\end{IEEEproof}
\vspace{2mm}

Property (b) implies that one can find the asymptotic rates
$ \sigma_{L}(\omega, \rho) $ and $ \sigma_{L}(\star, \rho) $ for any $ L $
if the rates for the left-most shift of $ L $ (i.e., $ L + 1 - \inf L $) are
known.
For example, the asymptotic rates of constant-weight $ (d, k) $-sequences
can be determined from those of $ (0, k-d) $-sequences by using \eqref{eq:sL2}
(see Remark \ref{rem:dk}).
Furthermore, by using this property, \eqref{eq:limitwr} and \eqref{eq:limitr},
one can obtain an explicit expression for the capacity for the case
$ L = \{ d+1, d+2, \ldots \} $ (corresponding to $ (d, \infty) $-sequences).
Namely, for every $ \rho \in \big[ 0, \frac{1}{d+1} \big] $ and
$ \omega \in \big[ \frac{\rho}{2} (d + 1) , 1 - \frac{\rho}{2} (d + 1) \big] $,
we have
\begin{align}
\nonumber
  \mysr_{\N + d}(\omega, \rho)
	 &= (1 - \omega - d \rho/2) H\!\left( \frac{ \rho/2 }{ 1 - \omega - d \rho/2 } \right)  \\
\label{eq:dinfty}
	    &\phantom{===} + (\omega - d \rho/2) H\!\left( \frac{ \rho/2 }{ \omega - d \rho/2 } \right) ,  \\
\label{eq:dinftyr}
  \mysr_{\N + d}(\star, \rho)
	 &= (1 - d \rho) H\!\left( \frac{ \rho }{ 1 - d \rho } \right) .
\end{align}

\subsection{Typical RLL Sequences}
\label{sec:typical}

As noted in the previous subsection, RLL sequences of weight $ \omega^\star n = \frac{1}{2} n $
and having $ \rho^\star n $ runs (where $ \rho^\star $ was defined in
\eqref{eq:rstar}) achieve the same asymptotic rate as RLL sequences without
additional weight or run constraints (see \eqref{eq:simwstar}), but with a
penalty of $ \sim\!\frac{\log n}{n} $ bits per symbol.
We wish to emphasize here that this penalty can be avoided if one relaxes the
conditions $ w = \frac{1}{2}n $, $ r = \rho^\star n $ to $ w \sim \frac{1}{2}n $,
$ r \sim \rho^\star n $.
To be more precise, there exists a \emph{sub-linear} function $ f(n) = o(n) $
such that, as $ n \to \infty $,
\begin{align}
\label{eq:typical2}
  \mySr_{L}(n)  \;\sim
	  \sum_{w, r \, : \, |w - \omega^\star n| + |r - \rho^\star n| \leqslant f(n) }  \mySr_{L}(n, w, r) .
\end{align}
Therefore, the number of RLL sequences whose weight and number of runs
satisfy%
\footnote{The condition $ |w - \omega^\star n| + |r - \rho^\star n| > f(n) $
can be replaced by different conditions stating that $ (w, r) $ is bounded
away from $ (\omega^\star n, \rho^\star n) $ (e.g., using $ \ell_2 $ instead
of $ \ell_1 $ distance), without affecting the validity of the statement.}
$ |w - \omega^\star n| + |r - \rho^\star n| > f(n) $ is asymptotically
negligible with respect to the number of all RLL sequences.
In fact, a stronger statement is true.

\begin{theorem}
Let $ \omega^\star = 1/2 $ and
$ \rho^\star = \left( \sum_{\ell \in L} \ell \lambda^\ell \right)^{\!-1} $,
where $ \lambda $ is the unique positive number satisfying
$ \sum_{\ell \in L} \lambda^\ell = 1 $.
There exists a sub-linear function $ f(n) = o(n) $ such that, as $ n \to \infty $,
\begin{align}
\label{eq:typical}
  \sum_{w, r \, : \, |w - \omega^\star n| + |r - \rho^\star n| > f(n) }
	 \mySr_{L}(n, w, r)
	  \;=\; {\mathcal O}\bigg(\frac{ \mySr_{L}(n) }{ n^{\log n} } \bigg) .
\end{align}
\end{theorem}
\begin{IEEEproof}
The proof is analogous to that of \cite[Lemmas 2 and 3]{kovacevic2}.
From the relation $ S_L(n) = \sum_{w, r} S_L(n, w, r) $, the pigeon-hole
principle, and the fact that the exponent $ \sigma_L(\omega, \rho) $ is
uniquely maximized at $ (\omega^\star, \rho^\star) $ (see Proposition
\ref{thm:properties}), we see that, for any given $ \epsilon > 0 $, the
number of RLL sequences satisfying
$ |w - \omega^\star n| + |r - \rho^\star n| > \epsilon n $ is exponential
with an exponent \emph{strictly smaller} than the exponent of $ S_L(n) $
(which is given in \eqref{eq:sigmamax}--\eqref{eq:lambda}).
More precisely, for every $ \epsilon > 0 $ there exists a (sufficiently small)
$ \gamma(\epsilon) > 0 $ such that, as $ n \to \infty $,
\begin{align}
\label{eq:CWasymp3}
  \sum_{w, r \, : \, |w - \omega^\star n| + |r - \rho^\star n| > \epsilon n }
	 \mySr_{L}(n, w, r)
	  \;\lesssim\;  \frac{ \mySr_{L}(n) }{ 2^{\gamma(\epsilon) n} } .
\end{align}
This further implies that, for every $ \epsilon > 0 $ and large enough $ n $,
\begin{align}
\label{eq:CWasymp4}
  \sum_{w, r \, : \, |w - \omega^\star n| + |r - \rho^\star n| > \epsilon n }
	 \mySr_{L}(n, w, r)
		\;<\;  \frac{ \mySr_{L}(n) }{ n^{\log n} } .
\end{align}
Let $ n_0(\epsilon) $ be the smallest positive integer such that \eqref{eq:CWasymp4}
holds for all $ n \geqslant n_0(\epsilon) $.
Now take an arbitrary sequence $ (\epsilon_i) $ satisfying
$ 1 = \epsilon_0 > \epsilon_1 > \epsilon_2 > \cdots $ and $ \lim_{i \to \infty} \epsilon_i = 0 $,
define the function
\begin{equation}
\label{eq:f}
  f(n)  =  \epsilon_i n ,  \qquad   n_0(\epsilon_i) \leqslant n < n_0(\epsilon_{i+1}) ,
\end{equation}
and note that \eqref{eq:CWasymp4} and \eqref{eq:f} imply \eqref{eq:typical}.
\end{IEEEproof}
\vspace{2mm}

It should be clear from the proof that the statement remains valid with
\emph{any} subexponential function placed instead of $ n^{\log n} $ (i.e.,
a function of the form $ 2^{g(n)} $ with $ g(n) = o(n) $), and therefore
also with $ o(\cdot) $ placed instead of $ {\mathcal O}(\cdot) $ on the
right-hand side of \eqref{eq:typical}.

We may paraphrase the relations \eqref{eq:typical2} and \eqref{eq:typical}
as follows: \emph{almost all} sequences in $ \mySSr_{L}(n) $ are balanced
(in the sense that their weight is $ \sim\!\frac{1}{2} n $) and have
$ \sim\!\rho^\star n $ runs.
For this reason, we say that $ \omega^\star = 1/2 $ is the ``typical''
relative weight, and $ \rho^\star $ the ``typical'' relative number of
runs in RLL sequences.
One application of the typicality property \eqref{eq:typical} that we
shall exhibit in Section \ref{sec:wpchannels} is that, in the asymptotic
analysis of the cardinality of optimal codes correcting a given number ($ t $)
of errors, due to the fact that $ n^t = o\big(n^{\log n}\big) $, one can
assume without loss of generality that all the codewords have
$ \sim\!\rho^\star n $ runs and are of weight $ \sim\!\frac{1}{2} n $.

RLL sequences exhibit typicality with respect to other parameters as well.
For example, almost all RLL sequences (in the sense analogous to \eqref{eq:typical})
contain $ \sim\! \beta^{\star}_{\ell} n $ runs of length $ \ell \in L $
(i.e., blocks $ 0^{\ell} $ and $ 1^\ell $), where
\begin{align}
\label{eq:betastar}
  \beta^{\star}_{\ell}
	 = \lambda^{\ell} \rho^\star
	 = \frac{ \lambda^{\ell} }{ \sum_{\ell' \in L} \ell' \lambda^{\ell'} } .
\end{align}
Note that $ \sum_{\ell\in L} \beta^{\star}_{\ell} = \rho^\star $
(see \eqref{eq:betastar} and \eqref{eq:lambda}),
as it should be because $ \rho^\star $ is the total typical number of
blocks in sequences from $ \mySSr_{L}(n) $.
The easiest way to derive \eqref{eq:betastar} is by using the probabilistic
interpretation from Section \ref{sec:probabilistic}: the probability of
length $ \ell $ under the \emph{optimal} distribution in \eqref{eq:maxent}
is $ P^\star_{\mathsf{L}}(\ell) = \lambda^\ell $, so the expected number
of blocks of length $ \ell $, in a sequence having $ \rho^\star n $ blocks
in total, is $ \lambda^\ell \rho^\star n $.
Based on the probabilistic interpretation it is easy to find other typical
values that may be of interest:
almost all sequences in $ \bs{S}_{L}(n) $ have
$ \sim\! P^\star_{\mathsf{L}}(\ell) P^\star_{\mathsf{L}}(\ell') \rho^\star n
= \lambda^{\ell + \ell'} \rho^\star n $ runs of length $ \ell $ that are
followed by a run of length $ \ell' $;
almost all sequences in $ \bs{S}_{L}(n) $ have
$ \sim\! \frac{1}{2} \sum_{\ell \in L, \ell \geqslant \underline{\ell}} 
  P^\star_{\mathsf{L}}(\ell) \rho^\star n
= \frac{1}{2} \sum_{\ell \in L, \ell \geqslant \underline{\ell}}
  \lambda^{\ell} \rho^\star n $
runs of \emph{zeros} of length greater than or equal to $ \underline{\ell} $
(and as many runs of \emph{ones} of length greater than or equal to $ \underline{\ell} $);
etc.

\begin{example}[Unconstrained case, continued]
\textnormal{
  For $ L = \N $, we have $ \lambda = 1/2 $, $ \rho^\star = 1/2 $,
$ P^\star_{\mathsf{L}}(\ell) = 2^{-\ell} $, and
$ \beta^{\star}_{\ell} = 2^{-\ell-1} $, for any $ \ell \in \N $.
}
\myqed
\end{example}

\subsection{Separate Constraints on the Runs of $ 0 $'s and $ 1 $'s}
\label{sec:different}

To conclude this section, we note that the presented results can be
generalized to the case where runs of $ 0 $'s and runs of $ 1 $'s
have possibly different and independent constraints \cite{kurmaev2}.
Namely, let $ S_{L_0, L_1}(n, w, r) $ denote the number of binary
sequences of length $ n $, weight $ w $, and containing $ r $ runs of
identical symbols, where the length of each run of $ 0 $'s (resp.\ $ 1 $'s)
belongs to $ L_0 \subseteq \N $ (resp.\ $ L_1 \subseteq \N $).
Then, for any $ (\omega, \rho) $ satisfying
\begin{subequations}
\label{eq:omegarhobounds01}
\begin{alignat}{2}
  \frac{\inf L_1}{\inf L_1 + \sup L_0}
	&<  \omega  &&<
	\frac{\sup L_1}{\inf L_0 + \sup L_1} ,  \\
	\max\! \left\{ \frac{2 \omega}{\sup L_1}, \, \frac{2 (1 - \omega)}{\sup L_0} \right\}
	&<  \rho  &&<
  \min\! \left\{ \frac{2 \omega}{\inf L_1}, \, \frac{2 (1 - \omega)}{\inf L_0} \right\} ,
\end{alignat}
\end{subequations}
we have, as $ n \to \infty $,
\begin{equation}
\label{eq:rllasymptotics01}
  \log S_{L_0, L_1}(n, \omega n, \rho n)
	 =  n \sigma_{L_0, L_1}(\omega, \rho) - \log n + {\mathcal O}(1) ,
\end{equation}
where
\begin{align}
\label{eq:rllcapacity01}
  &\sigma_{L_0, L_1}(\omega, \rho)  \\
\nonumber
	 &= - (1 - \omega) \log \alpha - \omega \log \beta
	    + \frac{\rho}{2} \log\!\left( \sum_{\ell \in L_0} \alpha^\ell \sum_{\ell \in L_1} \beta^{\ell} \right)
			\phantom{A}
\end{align}
and $ (\alpha, \beta) $ is the unique pair of positive real numbers satisfying
the equations
\begin{align}
\label{eq:alphabeta01}
  \sum_{\ell \in L_0} \left( \ell - \frac{2 (1 - \omega)}{\rho} \right) \alpha^\ell  =  0,  \quad
  \sum_{\ell \in L_1} \left( \ell - \frac{2 \omega}{\rho} \right) \beta^\ell  =  0 .
\end{align}
The remaining statements can be generalized in a similar way.
E.g., in Theorem \ref{thm:maxentrll}, $ P_{\mathsf{L}_0} $ and 
$ P_{\mathsf{L}_1} $ are probability distributions over $ L_0 $ and
$ L_1 $ respectively.

\section{Applications}
\label{sec:applications}

In this section we describe three communication scenarios in which asymptotic
properties of RLL sequences are used to derive the optimal information rates,
or bounds thereon.

\subsection{Weight-Preserving and Run-Preserving Channels}
\label{sec:wpchannels}

Several important classes of communication channels possess the property
of preserving certain qualities of input sequences, such as their weight
or the number of runs they contain.
Examples of weight-preserving channels include the bit-shift channel
\cite{shamai+zehavi}, which models a scenario in which an electric charge
may leak to one of the neighboring cells, resulting in the $ 1 $ bit that
it represents being ``shifted'' to the left or to the right, as well as
various other channels with permutation/reordering and timing errors.
An example of a channel that is run-preserving (i.e., that preserves the
number of runs in input sequences) is a sticky-insertion channel
\cite{dolecek+anantharam}, which models a scenario with synchronization
errors as a result of which one or more copies of a symbol may be inserted
next to the original symbol in the input sequence%
\footnote{As an aside related to sticky-insertion channels and RLL sequences,
we note that the set $ \bs{S}_L(n) $, for
$ L = \big\{ \frac{1}{h} \big( (h + 1)^j - 1 \big) : j \in \N \big\} $,
constitutes an optimal \emph{zero-error} code for the channel in which each
individual symbol may be copied at most $ h $ times \cite{kovacevic3}.}%
.

The weight (resp.\ number of runs) being unaffected by the channel means
that an optimal code can be represented as the union of optimal constant-weight
(resp.\ constant-number-of-runs) codes over all possible weights (resp.\
numbers of runs).
Furthermore, by the discussion in Section \ref{sec:typical}, for the purpose
of determining the asymptotic behavior of optimal codes one may assume with
no loss in generality that all the codewords are typical.
These facts together considerably simplify the analysis (see, e.g., \cite{kovacevic2}).
We illustrate this below on the example of a deletion channel with RLL inputs.

\subsubsection*{RLL Insertion/Deletion Channels}

Consider the deletion channel with inputs from $ \bs{S}_L(n) $, where
$ L = \{d+1, \ldots, k+1\} $.
If the maximum possible number of deleted symbols is $ t \leqslant d $,
then the channel is run-preserving as it cannot delete entire runs of
identical symbols.
This fact was used in \cite{paluncic} to devise a construction of codes
correcting $ t \leqslant d $ deletions.
In the following theorem we give the asymptotic scaling of the construction
from \cite{paluncic}, thus obtaining a lower bound on the cardinality of
optimal $ t $-deletion-correcting RLL codes, and we also derive an upper
bound on this quantity.
For $ t = 1 $ and $ k = \infty $, the bounds coincide and give the exact
asymptotic scaling of the cardinality of optimal single-deletion-correcting
codes.
We emphasize that the presented bounds are also valid for $ t > d $ in
the restricted (run-preserving) deletion channel in which an additional
assumption is adopted that the channel cannot delete all bits from a given
run, i.e., that it cannot delete entire runs.

Denote the size of an optimal $ t $-deletion-correcting code in the space
$ \bs{S}_L(n) $ (resp.\ $ \bs{S}_{L}(n, \star, r ) $) by $ D_{L}(n; t) $
(resp.\ $ D_{L}(n, r; t) $).
Thus $ D_{L}(n, r; t) $ is the size of an optimal code having the property
that all the codewords contain exactly $ r $ runs.
If the channel is run-preserving, it must hold that
$ D_{L}(n; t) = \sum_{r} D_{L}(n, r; t) $.

\begin{theorem}
\label{thm:bounds_del}
Fix $ t, d, k $ with $ 0 \leqslant t \leqslant d < k \leqslant \infty $,
and denote $ L = \{d+1, \ldots, k+1\} $.
As $ n \to \infty $,
\begin{align}
\label{eq:bounds}
   \frac{ S_{L}(n) }{ ( \rho^\star n )^{t} }
	 \,\lesssim\,
   D_{L}(n; t)
	 \,\lesssim\,
	 \frac{ S_{L}(n) }{ ( \rho^\star n )^{t} } \frac{ \lambda^t t! }{ ( 1 - \lambda^{d+1} )^t }  ,
\end{align}
where $ \rho^\star = \big( \sum_{\ell=d+1}^{k+1} \ell \lambda^\ell \big)^{\!-1} $,
and $ \lambda $ is the unique positive number satisfying
$ \sum_{\ell=d+1}^{k+1} \lambda^\ell = 1 $.

For $ k = \infty $ and $ t = 1 $,
\begin{align}
\label{eq:exact}
   D_{L}(n; 1)
	 \,\sim\,
	 \frac{ S_{L}(n) }{ n } \left( d + \frac{1}{1 - \lambda} \right) .
\end{align}
\end{theorem}
\begin{IEEEproof}
The lower bound in \eqref{eq:bounds} can be obtained either by using
the construction from \cite{paluncic}, or by generalizing the construction
from \cite{levenshtein} to $ (d,k) $-constrained inputs and using \cite{bose+chowla}
(see \cite{kovacevic+tan_clet} for details).
Either construction implies that, for fixed $ t $, as $ n \to \infty $ and
$ r \sim \rho n $, $ D_{L}(n, r; t) \gtrsim \frac{1}{r^t} S_{L}(n, \star, r) $.
From this and the typicality properties discussed in Section~\ref{sec:typical},
we get
\begin{subequations}
\label{eq:D}
\begin{align}
  D_{L}(n; t)
    \label{eq:aa}
    &=        \sum_{r=0}^n  D_{L}(n, r; t)   \\
    &\geqslant     \sum_{r = \rho^\star n - f(n)}^{\rho^\star n + f(n)}  D_{L}(n, r; t)   \\
		\label{eq:1}
		&\gtrsim  \frac{ 1 }{ \big(\rho^\star n + f(n)\big)^t }
		           \sum\limits_{r = \rho^\star n - f(n)}^{\rho^\star n + f(n)}
		            S_{L}(n, \star, r)   \\
    \label{eq:2}
		&\sim     \frac{ S_{L}(n) }{ (\rho^\star n)^t } .
\end{align}
\end{subequations}

The upper bound in \eqref{eq:bounds} is obtained by a packing argument.
Let $ \bs{D} \subseteq \bs{S}_{L}(n) $ be an optimal code correcting
$ t $ deletions, $ |\bs{D}| = D_{L}(n; t) $.
Consider a \emph{typical} codeword $ \myx \in \bs{D} $ having
$ r \sim \rho^\star n $ runs and
$ b \sim \beta_{d+1}^\star n = \lambda^{d+1} \rho^\star n $ runs of
length $ d + 1 $ (see \eqref{eq:betastar}).
Consider the patterns of $ t $ deletions such that at most one deletion
occurs in each run of length $ \geqslant\! d + 2 $ in $ \myx $.
The number of different sequences that can be produced after $ \myx $
is impaired by such a pattern is
$ \binom{r-b}{t} \sim \frac{1}{t!} (r - b)^t \sim
  \frac{1}{t!} \big( (1 - \lambda^{d+1}) \rho^\star n \big)^t $,
and all such sequences live in $ \bs{S}_{L}(n-t) $ (since deletions
occur in runs of length $ \geqslant\! d + 2 $, the resulting runs
are of length $ \geqslant\! d + 1 $).
Now, since $ \bs{D} $ is assumed to correct $ t $ deletions, the sets of
output sequences that can be obtained in this way from any two distinct
codewords must be disjoint.
Therefore,
\begin{align}
\label{eq:upper}
  D_{L}(n; t) \cdot \frac{ \left( \left( 1 - \lambda^{d+1} \right) \rho^\star n \right)^t }{ t! }
    \, \lesssim \,  S_{L}(n-t)  \,\sim\,  \lambda^{t} S_{L}(n) ,
\end{align}
which proves the stated upper bound.

When $ k = \infty $, we have
$ \sum_{\ell=d+1}^\infty \lambda^\ell = \frac{\lambda^{d+1}}{1 - \lambda} = 1 $
and $ {\rho^\star}^{-1} = \sum_{\ell=d+1}^\infty \ell \lambda^\ell = d + \frac{1}{1-\lambda} $.
If, in addition, $ t = 1 $, the two bounds from \eqref{eq:bounds} coincide
and give \eqref{eq:exact}.
\end{IEEEproof}
\vspace{2mm}

The constant factor in the upper bound in \eqref{eq:bounds} can be
improved for $ t \geqslant 2 $ by using the fact that a deletion-correcting
code is also an insertion-deletion-correcting code \cite{paluncic}
(see \cite[Theorem 4]{kovacevic+tan_clet}), but we satisfy ourselves here
with the fact that the two bounds have the same asymptotic scaling, namely
\begin{equation}
\begin{aligned}
  \log D_{L}(n; t)
	 &= \log S_{L}(n)  - t \log n + {\mathcal O}(1)  \\
	 &= n \log \lambda^{-1}  - t \log n + {\mathcal O}(1) ,
\end{aligned}
\end{equation}
where $ \lambda $ is the unique positive number satisfying
$ \sum_{\ell=d+1}^{k+1} \lambda^\ell = 1 $.
In words, redundancy of $ \sim\! t \log n $ bits is necessary and sufficient
for achieving communication resilient to $ t \leqslant d $ deletions.
We emphasize again that the same holds for $ t > d $ in the \emph{run-preserving}
deletion channel.
For example, in the unconstrained ($ L = \N $) run-preserving deletion channel
we have, for any $ t \geqslant 1 $,
\begin{align}
\label{eq:bounds1}
   \frac{ 2^{n+t} }{ n^{t} }
	 \,\lesssim\,
   D_{\N}(n; t)
	 \,\lesssim\,
	 \frac{ 2^{n+t} }{ n^{t} } t!  ,
\end{align}
and hence, for $ t = 1 $, $ D_{\N}(n; 1) \sim \frac{ 2^{n+1} }{ n } $.
For comparison, recall that the size of an optimal code correcting a single
deletion (with no run-preserving assumption) scales as $ \frac{ 2^{n} }{ n } $
\cite{levenshtein2}.

To conclude this subsection, let us reiterate the main idea behind the above
proof: in addition to using the fact that the channel is run-preserving, we
have restricted the analysis to typical codewords only, a restriction which
simplified the proof while not incurring any loss in generality.
In particular, we have exploited the fact that the sets of output sequences
are of the same size for all codewords having the same number of runs $ r $,
and that we may freely take $ r \sim \rho^\star n $.

\subsection{Channels with Sparse Error Patterns}
\label{sec:sparse}

Consider the binary symmetric channel (BSC), in which an output sequence
is produced by flipping some of the symbols/bits in the input sequence,
but with an additional constraint that, in any contiguous segment of $ d + 1 $
bits, only one bit may be flipped.
This channel models a communication scenario in which the error patterns
are ``sparse'' in the sense that two input bits may be affected by noise
only if they are sufficiently far apart.
Denoting the input, output, and noise sequences by
$ \bs{x}, \bs{y}, \bs{e} \in \{0, 1\}^n $, respectively, we can write
$ \bs{y} = \bs{x} \oplus \bs{e} $, where the requirement that the error
patterns are sparse is equivalent to saying that any two $ 1 $'s in the
sequence $ \bs{e} $ are separated by at least $ d $ $ 0 $'s.
In other words, the sequence $ \bs{e} $ is a $ (d, \infty) $-sequence%
\footnote{Note that not all run-lengths of zeros in $ \bs{e} $ are constrained
to the set $ \{ d, d + 1, \ldots \} $ -- the first (if $ e_1 = 0 $) and the
last (if $ e_n = 0 $) run can be shorter. Since allowing the first and/or
the last run to violate the given constraints does not affect the asymptotic
rate of constrained sequences, we do not pay particular attention to these
boundary conditions.}
(see Remark \ref{rem:dk}).
Hence, in this scenario, the noise sequences, rather than the information
sequences, are constrained.

Let $ V_d(n, r) $ be the number of different output sequences that can
be produced by a given input sequence $ \bs{x} \in \{0, 1\}^n $, assuming
that at most $ r $ errors have occurred in the channel, and define
\begin{equation}
  v_d(\rho) = \lim_{n \to \infty} \frac{1}{n} \log V_d(n, \rho n) ,
\end{equation}
for $ \rho \in \big[ 0, \frac{1}{d+1} \big] $.

\begin{lemma}
\label{thm:vol}
For any $ d \in \{0, 1, \ldots\} $,
\begin{equation}
\label{eq:vol}
  v_d(\rho) =
	\begin{cases}
	 (1 - d \rho) H\!\left( \frac{\rho}{1 - d \rho} \right) ,
	   &  0 \leqslant \rho < \frac{ 1 - \lambda }{ 1 + (1 - \lambda) d }  \\
	 - \log \lambda ,
	   &  \frac{ 1 - \lambda }{ 1 + (1 - \lambda) d } \leqslant \rho \leqslant \frac{1}{d+1} ,
	\end{cases}
\end{equation}
where $ \lambda $ is the unique positive solution of the equation
$ \lambda^{d+1} + \lambda - 1 = 0 $.
\end{lemma}
\begin{IEEEproof}
$ V_d(n, r) $ is equal to the number of different noise sequences of length
$ n $ and Hamming weight $ \leqslant\! r $.
Since the noise sequences are $ (d, \infty) $-sequences (with boundary
conditions different from those assumed in Remark \ref{rem:dk}), and since
the boundary conditions do not affect the asymptotic rate, we conclude that
$ \log V_d(n, \rho n)  \sim  \log \sum_{t \leqslant \rho n} S_{\N+d}(n, \star, t) $
(see Remark \ref{rem:dk} and the first paragraph of Section \ref{sec:ronly}).
We then have
\begin{subequations}
\label{eq:Sbexp}
\begin{align}
  v_d(\rho)
	 &=  \lim_{n\to\infty}  \frac{1}{n} \log \sum_{t = 0}^{\rho n} \mySr_{\N+d}(n, \star, t)  \\
\label{eq:pigeon}
	 &=  \lim_{n\to\infty}  \frac{1}{n} \log \max_{ 0 \leqslant t \leqslant \rho n} \mySr_{\N+d}(n, \star, t)  \\
\label{eq:s1}
	 &=  \max_{\tau \in \left[ 0 ,\, \rho \right]} \mysr_{\N+d}(\star, \tau)  \\
\label{eq:temp}
	 &=  \begin{cases}
	       \mysr_{\N+d}(\star, \rho)  , &  0 \leqslant \rho < \rho^\star  \\
				 \mysr_{\N+d}(\star, \star) , &  \rho^\star \leqslant \rho \leqslant \frac{1}{d+1}
	     \end{cases} ,
\end{align}
\end{subequations}
where \eqref{eq:pigeon} follows from the pigeon-hole principle;
\eqref{eq:s1} follows from the definition of $ \sigma_L(\star, \rho) $
(see \eqref{eq:Sexp} and \eqref{eq:expr}); and \eqref{eq:temp} follows
from the fact that the mapping $ \rho \mapsto \sigma_L(\star, \rho) $
is concave and maximized at $ \rho^\star $ (see Proposition \ref{thm:properties}(d)).
Now recall from \eqref{eq:dinftyr} that the function $ \mysr_{L}(\star, \rho) $
has an explicit form for $ L = \N + d $, namely
$ \mysr_{\N+d}(\star, \rho) = (1 - d \rho) H\big( \frac{ \rho }{ 1 - d \rho } \big) $,
simplify \eqref{eq:rstar} to
$ \rho^\star = \left( \sum_{\ell = d}^\infty \ell \lambda^\ell \right)^{\!-1}
  = \frac{1 - \lambda}{1 + (1 - \lambda) d} $, 
where $ \lambda $ is the unique positive number satisfying
$ \sum_{\ell = d}^\infty \lambda^\ell = \frac{\lambda^d}{1 - \lambda} = 1 $,
and recall from \eqref{eq:sigmamax} that $ \sigma_{\N+d}(\star, \star) = - \log \lambda $.
\end{IEEEproof}
\vspace{2mm}

Denote the maximum cardinality of a code of length $ n $ correcting $ r $
errors (as defined above) by $ M_d(n, r) $.
Then, by the sphere packing argument, we must have
$ M_d(n, r) \leqslant \frac{2^n}{V_d(n, r)} $.
Consequently, the maximum asymptotic rate of codes correcting a \emph{fraction}
$ \rho $ of errors is upper bounded by
\begin{equation}
\label{eq:SPm}
  \lim_{n \to \infty} \frac{1}{n} \log M_d(n, \rho n)  \leqslant  1 - v_{d}(\rho) .
\end{equation}
For $ d = 0 $ (unconstrained BSC), \eqref{eq:vol} reads
\begin{equation}
  v_0(\rho) =
	\begin{cases}
	 H(\rho) ,  &  0 \leqslant \rho < \frac{1}{2}  \\
	    1 ,     &  \frac{1}{2} \leqslant \rho \leqslant 1 ,
	\end{cases}
\end{equation}
and hence \eqref{eq:SPm} reduces to the classical
sphere packing bound.

The maximum asymptotic rate of codes correcting a given fraction $ \rho $
of errors is a quantity of fundamental importance in coding theory, the
exact value of which is still unknown for any non-trivial $ \rho $.
Upper bounds better than the sphere-packing bound are known \cite{vanlint},
and the best known lower bound is the Gilbert--Varshamov bound.
Whether arguments used for deriving these bounds can be applied in our
scenario with sparse error patterns will be investigated elsewhere;
the derivation of \eqref{eq:SPm} was given here only as an illustration.

\subsection{Sub-block Constrained Sequences}

We conclude this section by presenting another type of constrained sequences.
Their analysis does not represent an application of the above results per se,
but rather an illustration that a wider class of problems can be analyzed by
using similar methods.

For fixed $ \ell_\myb, w_\myb $ with $ w_\myb \leqslant \ell_\myb $, define
$ \B = \big\{ \myx \in \{0,1\}^{\ell_\myb} : \weight(\myx) \geqslant w_\myb \big\} $,
and consider the set of binary sequences of length $ n $ and weight $ w $ obtained
by concatenating blocks from $ \B $, namely
\begin{equation}
\label{eq:subblock}
  \bs{S}^\textnormal{sec}(n, w)
	 = \left\{ \myx \in \B^{\frac{n}{\ell_\myb}} : \weight(\myx) = w \right\} ,
\end{equation}
where it is assumed that $ n/\ell_\myb $ is an integer.
(We omit the parameters $ \ell_\myb, w_\myb $ from the notation for reasons of
simplicity.)
Denote also, as before, $ S^\textnormal{sec}(n, w) = \big| \bs{S}^\textnormal{sec}(n, w) \big| $.

Sequences built from blocks of length $ \ell_\myb $ and weight $ \geqslant\! w_\myb $
are called sub-block energy constrained (SEC) sequences, and they have been
studied in the context of simultaneous information and energy transfer, as
well as in several other applications \cite{tandon2, tandon}.
From the definition \eqref{eq:subblock} it follows that the quantity
$ S^\textnormal{sec}(n, w) $ obeys the recurrence relation
\begin{align}
  S^\textnormal{sec}(n, w)
	 = \sum_{j=w_\myb}^{\ell_\myb} \binom{\ell_\myb}{j}  S^\textnormal{sec}(n - \ell_\myb, w - j) ,
\end{align}
the characteristic equation of which is
\begin{align}
  1 - x^{\ell_\myb} \sum_{j=w_\myb}^{\ell_\myb} \binom{\ell_\myb}{j}  y^j  =  0 .
\end{align}
Applying \cite[Theorem 1.3]{pemantle+wilson1}, we obtain the following
characterization of the asymptotic rate of \emph{constant-weight} SEC
sequences.

\begin{theorem}
\label{thm:subblock}
Fix $ \ell_\myb, w_\myb $, and $ \omega \in \big( \frac{w_\myb}{\ell_\myb}, 1 \big) $.
As%
\footnote{The condition $ n \to \infty $ is here understood over
$ \ell_\myb \N = \{ \ell_\myb m : m \in \N \} $.}
$ n \to \infty $,
\begin{equation}
  \log S^\textnormal{sec}(n, \omega n)
	 = n \sigma^\textnormal{sec}(\omega) - \frac{1}{2} \log n + \mathcal{O}(n) ,
\end{equation}
where
\begin{equation}
  \sigma^\textnormal{sec}(\omega)
	 = - \omega \log \beta + \frac{1}{\ell_\myb} \log \sum_{j=w_\myb}^{\ell_\myb} \binom{\ell_\myb}{j} \beta^j ,
\end{equation}
and $ \beta $ is the unique positive number satisfying
$ \sum_{j=w_\myb}^{\ell_\myb} \binom{\ell_\myb}{j} (j - \omega \ell_\myb) \beta^j = 0 $.
\end{theorem}

We also set
$ \sigma^\textnormal{sec}(w_\myb/\ell_\myb)
  = \sigma^\textnormal{sec}(1) = 0 $.
The exponent $ \sigma^\textnormal{sec}(\omega) $ is a continuous,
strictly concave function of $ \omega \in \big[ \frac{w_\myb}{\ell_\myb}, 1 \big] $,
maximized at
\begin{equation}
  \omega^\star = \frac{ \sum_{j=w_\myb}^{\ell_\myb} j \binom{\ell_\myb}{j} }{ \ell_\myb \sum_{j=w_\myb}^{\ell_\myb} \binom{\ell_\myb}{j} } ,
\end{equation}
with the maximum being equal to
\begin{equation}
  \sigma^\textnormal{sec}(\omega^\star)
	 = \frac{1}{\ell_\myb} \log \sum_{j=w_\myb}^{\ell_\myb} \binom{\ell_\myb}{j} .
\end{equation}
The last equality can also be obtained directly as
$ \sigma^\textnormal{sec}(\omega^\star) $ must be equal to the exponential
growth rate of SEC sequences without the constant-weight requirement, namely
$ \lim_{n \to \infty} \frac{1}{n} \log \big| \B^{\frac{n}{\ell_\myb}} \big| $.
 
The exponential growth rate of the quantity
$ \sum_{w\geqslant \omega n} S^\textnormal{sec}(n, w) $ (bounded-weight SEC
sequences), which is also naturally of interest in this context, can be found
from the above by using the fact that $ \sigma^\textnormal{sec}(\omega) $ is
uniquely maximized at $ \omega^\star $, similarly as in \eqref{eq:Sbexp}.

\section{Constant-Manhattan-Weight Sequences}
\label{sec:manhattan}

There are many scenarios in coding theory where other definitions
of sequence weight are more appropriate than the Hamming weight, an
important example being the $ \ell_1 $ (or Manhattan) weight.
We wish to point out here that $ q $-ary sequences with fixed $ \ell_1 $
weight are, in a sense, dual to sequences with constrained runs of zeros
and that, consequently, their properties can be inferred from the results
of Section \ref{sec:rll}.

Denote $ \A_q = \{0, 1, \ldots, q-1\} $,
\begin{equation}
  \bs{M}_{q}(n, w)
	 = \left\{ \bs{x} \in \A_q^n : \sum_{i=1}^n x_i = w \right\} ,
\end{equation}
and $ M_{q}(n, w) = \big| \bs{M}_{q}(n, w) \big| $.
Interpreting the sequences in $ \bs{M}_{q}(n, w) $ as $ n $-part compositions
of the number $ w $, with parts restricted to $ \A_q $, we can write%
\footnote{Recall that $ S_{L}(w, \star, n) = 2 C_L(w, n) $, but we cannot
write here $ M_{q}(n, w) = \frac{1}{2} S_{\A_q}(w, \star, n) $ because
$ 0 \in \A_q $, and in the definition of $ \myS_{L}(\cdot, \cdot, \cdot) $ we
have assumed that $ 0 \notin L $ (it is meaningless to speak of runs of length
$ 0 $ in this context).
However, the fact that $ 0 \in \A_q $ does not affect the analysis \cite{pemantle+wilson}.
We note that integer compositions in which $ 0 $'s are valid parts are called
\emph{weak} compositions.}
$ M_{q}(n, w) = C_{\A_q}(w, n) $.
We can then deduce the following claim from \cite[Theorem 12.2.2]{pemantle+wilson},
in the same way as, e.g., Theorem \ref{thm:asymp}.

\begin{theorem}
Fix $ \omega \in \big( 0 , q-1 \big) $.
As $ n \to \infty $,
\begin{align}
\label{eq:Sexpm}
  \log M_{q}(n, \omega n)  =  n\, \mu_{q}(\omega)  -  \frac{1}{2} \log n  +  {\mathcal O}(1) ,
\end{align}
where
\begin{align}
\label{eq:expm}
  \mu_{q}(\omega)  =  - \omega \log \alpha + \log\sum_{i=0}^{q-1} \alpha^{i} ,
\end{align}
and $ \alpha $ is the unique positive real number satisfying%
\begin{equation}
\label{eq:rho3}
  \sum_{i=0}^{q-1} (i - \omega) \alpha^{i} = 0 .
\end{equation}
\end{theorem}

By continuous extension, we set $ \mu_{q}(0) = \mu_{q}(q-1) = 0 $.

\begin{proposition}
The mapping $ \omega \mapsto \mu_{q}(\omega) $ is continuous, symmetric,
and strictly concave on the interval $ \omega \in \big[ 0 , q-1 \big] $.
It is maximized at $ \omega^\star = \frac{q-1}{2} $, and the maximum is
equal to
\begin{equation}
  \mu_{q}(\omega^\star) =  \log q .
\end{equation}
\end{proposition}

The ``typical'' Manhattan weight of a $ q $-ary sequence of length
$ n \to \infty $ is $ \sim\! \frac{q-1}{2} n $.
The typical number of occurrences of each of the symbols $ a \in \A_q $ in
such a sequence is $ \sim\! \frac{1}{q} n $;
the typical number of non-zero symbols that are followed by a zero
symbol is $ \sim\! \frac{q-1}{q} \frac{1}{q} n = \frac{q-1}{q^2} n $; etc.

Note that we are analyzing here sequences with fixed Manhattan weight but
with no RLL constraints.
Deriving the asymptotics for the case when both these kinds of constraints
are taken into account is also an interesting problem and will be investigated
elsewhere.

\subsection*{Application: Molecular Timing Channels}

In this subsection we present an example of a Manhattan-weight-preserving
channel (that is not Hamming-weight-preserving) and an application of the
typicality properties just mentioned.

Consider a communication channel described by the following assumptions:
\begin{inparaenum}
\item[1)]
an input sequence $ \myx = x_1 \cdots x_n \in \A_q^n $, $ \sum_{i=1}^n x_i = w $,
is thought of as describing $ w $ identical particles transmitted over $ n $
time slots, where $ x_i $ particles are transmitted in the $ i $'th slot;
\item[2)]
all particles reach the receiver, but their relative positions at the receiving
side may differ from the ones they had at the moment of transmission, i.e., the
delay of each of the particles may vary with respect to the expected/mean delay.
\end{inparaenum}
This channel is meant to model a type of molecular communication systems
\cite{cui+eckford, kovacevic+popovski}, and it can also be seen as a
generalization of the bit-shift channel to $ q $-ary alphabets.

Note that every input sequence $ \bs{x} $ can be uniquely represented by
another sequence $ \bs{\tilde{x}} \in \Z^w $ satisfying
$ 1 \leqslant \tilde{x}_1 \leqslant \cdots \leqslant \tilde{x}_w \leqslant n $,
where $ \tilde{x}_j $ represents the time slot in which the $ j $'th particle
is sent.
For example, if $ \myx = (2, 1, 0, 2) $, then $ \bs{\tilde{x}} = (1, 1, 2, 4, 4) $.
A shift by one position of the $ j $'th particle in $ \myx $ (i.e., its delay
being one slot less or one slot greater than the mean delay) corresponds to a
$ \pm 1 $ error on the $ j $'th location in $ \bs{\tilde{x}} $.
We say that $ t $ particle shifts (or timing errors) have occurred in the
channel if $ \left\| \bs{\tilde{x}} - \bs{\tilde{y}} \right\|_1
  = \sum_{i=1}^w |\tilde{x}_i - \tilde{y}_i| = t $, where $ \bs{\tilde{y}} $
is the integer representation of the corresponding received sequence.
This representation of input and output sequences implies that a code
correcting $ t $ particle shifts can be constructed from a code in
$ \mathbb{Z}^w $ of minimum distance $ \geqslant\! 2t+1 $ under the Manhattan
metric \cite{kovacevic2}.

Let $ T_{q}(n; t) $ (resp.\ $ T_{q}(n, w; t) $) denote the cardinality of
an optimal code in $ \A_q^n $ (resp.\ $ \bs{M}_{q}(n, w) $) correcting
$ t $ timing errors.
Hence, $ T_{q}(n, w; t) $ is the cardinality of an optimal code all of
whose codewords have Manhattan weight $ w $.
Since the channel is Manhattan-weight-preserving, we must have
$ T_{q}(n; t) = \sum_{w} T_{q}(n, w; t) $.

\begin{theorem}
\label{thm:boundsm}
Fix $ q \geqslant 2 $ and $ t \geqslant 1 $.
As $ n \to \infty $,
\begin{align}
\label{eq:boundsm}
	 \frac{ q^n }{ n^{t} } \frac{ 2^t c(t) }{ (q - 1)^{t} }
	 \,\lesssim\,
	 T_{q}(n; t)
	 \,\lesssim\,
   \frac{ q^n }{ n^{t} } \frac{ q^{2t} t! }{ 2^t (q-1)^{2t} }   ,
\end{align}
where $ c(1) = \frac{1}{2} $, $ c(2) = \frac{1}{4} $, and $ c(t) = \frac{1}{2t+1} $
for $ t \geqslant 3 $.
\end{theorem}
\begin{IEEEproof}
As in the proof Theorem \ref{thm:bounds_del} (Section \ref{sec:wpchannels}),
the idea is to use the fact that the channel is Manhattan-weight-preserving,
and to restrict the analysis to typical codewords only (which incurs no loss
in generality).

As explained above, by using a different (integer) representation of input
and output sequences, a code correcting $ t $ shifts can be constructed
from a code of minimum distance $ \geqslant\! 2t+1 $ under the Manhattan
metric.
Since the best known packing of balls of radius $ t $ (in the Manhattan
metric) in $ \Z^w $ has center-density $ \gtrsim\! \frac{c(t)}{w^t} $
\cite[Lemma 10]{kovacevic2}, it follows that
$ T_{q}(n, w; t) \gtrsim M_q(n,w) \frac{c(t)}{w^t} $.
By using the typicality argument and letting
$ w \sim \omega^\star n = \frac{q-1}{2} n $, the lower bound in \eqref{eq:boundsm}
follows by the same reasoning as in \eqref{eq:D}.

Let us now establish the upper bound.
Let $ \C \subseteq \A_q^n $ be an optimal code correcting $ t $ timing
errors, $ |\C| = T_{q}(n; t) $, and consider a codeword $ \myx \in \C $.
Every pattern of $ t $ shifts that can impair $ \myx $ in the channel
consists of $ s $ right-shifts and $ t - s $ left-shifts, for some
$ s \in \{0, 1, \ldots, t\} $.
We consider only shifts (from non-empty slots) to slots which are not
``full'', i.e., which have $ < q - 1 $ particles, so that the resulting
sequence at the channel output also belongs to $ \A_q^n $.
The typical number of non-empty slots with a ``non-full'' slot immediately
to their right is easily shown to be
$ \gamma^\star n = \big( \frac{q-1}{q} \big)^{\!2} n $, and the same is
true for the number of non-empty slots with a ``non-full'' slot immediately
to their left.
Based on this, one concludes that the number of different sequences in
$ \A_q^n $ that can be obtained after $ \myx $ is impaired by $ t $ shifts
is (asymptotically) lower-bounded by%
\begin{equation}
\label{eq:shifts}
  \gtrsim\,
   \sum_{s=0}^t
	  \binom{ \gamma^\star n  }{ s }
    \binom{ \gamma^\star n }{ t - s }
		\,\sim\,
		(\gamma^\star n)^t \frac{ 2^t }{ t! } .
\end{equation}
Since $ \C $ corrects $ t $ shifts by assumption, we must have
\begin{align}
  T_{q}(n; t)  \cdot
     \frac{ 2^t }{ t! }  \left( \frac{q-1}{q} \right)^{\!2t}  n^t
   \,\lesssim\,  q^n  ,
\end{align}
which is what we needed to show.
\end{IEEEproof}
\vspace{2mm}

Consequently, as $ n \to \infty $,
\begin{equation}
  \log T_{q}(n; t)  =  n \log q - t \log n  +  {\mathcal O}(1) .
\end{equation}

\vspace{6mm}
\balance

\end{document}